\RequirePackage{rotating}
\documentclass[fleqn,usenatbib]{mnras}

\usepackage{newtxtext,newtxmath}

\usepackage[T1]{fontenc}

\DeclareRobustCommand{\VAN}[3]{#2}
\let\VANthebibliography\thebibliography
\def\thebibliography{\DeclareRobustCommand{\VAN}[3]{##3}\VANthebibliography}

\usepackage{graphicx}	
\usepackage{amsmath}	
\usepackage{amssymb}	

\usepackage{longtable}	
\usepackage{rotating}
\usepackage[normalem]{ulem} 
\usepackage{algorithm}  
\usepackage{algpseudocode}
\usepackage{fontawesome} 
\usepackage{pdflscape} 
\usepackage{subcaption}
\usepackage{caption}
\usepackage{soul,xcolor} 
\setstcolor{red}  

\title[AUDS100]{The Arecibo Ultra-Deep Survey}

\author[Hongwei Xi, Lister Staveley-Smith, Bi-Qing For et al.]
{
	Hongwei Xi$^{1,2,3,4}$, Lister Staveley-Smith$^{2,4}$\thanks{Contact e-mail: \href{mailto:lister.staveley-smith@uwa.edu.au}{lister.staveley-smith@uwa.edu.au}}, Bi-Qing For$^{2,4}$, Wolfram Freudling$^5$, 
	\newauthor
	Martin Zwaan$^5$, Laura Hoppmann$^{2}$, Fu-Heng Liang$^{2,6}$, and Bo Peng$^1$\\
	$^1$ CAS Key Laboratory of FAST, National Astronomical Observatories, Chinese Academy of Sciences, Datun Rd., Chaoyang District, Beijing 100101, China\\
	$^2$ International Centre for Radio Astronomy Research (ICRAR), University of Western Australia, 35 Stirling Hwy, Crawley, WA 6009, Australia\\
	$^3$ University of Chinese Academy of Sciences, Beijing 100049, China\\
	$^4$ ARC Centre of Excellence for All Sky Astrophysics in 3 Dimensions (ASTRO 3D), Australia\\
	$^5$ European Southern Observatory, Karl-Schwarzschildstrasse 2, D-85748 Garching bei M\"unchen, Germany\\
	$^6$ Department of Astronomy, Tsinghua University, Beijing 100084, China\\
}

\date{Accepted XXX. Received YYY; in original form ZZZ}

\pubyear{2015}

\graphicspath{{./Figures/}}

\begin{document}
\label{firstpage}
\pagerange{\pageref{firstpage}--\pageref{lastpage}}
\maketitle

\begin{abstract}

    The Arecibo Ultra Deep Survey (AUDS) is a blind HI survey aimed at detecting galaxies beyond the local Universe in the 21-cm emission line of neutral hydrogen (HI). The Arecibo $L$-band Feed Array (ALFA) was used to image an area of 1.35~deg$^2$ to a redshift depth of 0.16, using a total on-source integration time of over 700 hours. The long integration time and small observation area makes it one of the most sensitive HI surveys, with a noise level of $\sim 75$~$\mu$Jy per  21.4~kHz (equivalent to 4.5~km~s$^{-1}$ at redshift $z=0$). We detect 247 galaxies in the survey, more than doubling the number already detected in AUDS60. The mass range of detected galaxies is $\log(M_{\rm HI}~[h_{70}^{-2}{\rm M}_\odot]) = 6.32 - 10.76$. A modified maximum likelihood method is employed to construct an HI mass function (HIMF). The best fitting Schechter parameters are: low-mass slope $\alpha = -1.37 \pm 0.05$, characteristic mass $\log(M^*~[h_{70}^{-2}{\rm M}_\odot]) = 10.15 \pm 0.09$, and density $\Phi_* = (2.41 \pm 0.57) \times 10^{-3} h_{70}^3$~Mpc$^{-3}$~dex$^{-1}$. The sample was divided into low and high redshift bins to investigate the evolution of the HIMF. No change in low-mass slope $\alpha$ was measured, but an increased characteristic mass $M^*$, was noted in the higher-redshift sample. Using Sloan Digital Sky Survey (SDSS) data to define relative galaxy number density, the dependence of the HIMF with environment was also investigated in the two AUDS regions. We find no significant variation in $\alpha$ or $M^*$. In the surveyed region, we measured a cosmic HI density $\Omega_{\rm HI} = (3.55 \pm 0.30) \times 10^{-4} h_{70}^{-1}$. There appears to be no evolutionary trend in $\Omega_{\rm HI}$ above $2\sigma$ significance between redshifts of 0 and 0.16.
    
\end{abstract}




\section{Introduction}\label{intro}

	Neutral hydrogen (HI) is the most abundant component in the interstellar medium of late-type galaxies. It is an important reservoir for feeding gas into molecular clouds and star-forming regions \citep{2008A&ARv..15..189S}. The distribution function of the HI masses of galaxies (the HI mass function, HIMF) and the cosmic HI density ($\Omega_{\rm HI}$) are crucial measurements in characterising galaxies at different redshifts, and allow quantitative comparisons with theoretical models and simulations. Observations also show that HI has an extended distribution in galaxies relative to the stellar content and other tracers, therefore better filling the dark matter halo, better tracing the outflow and inflow of material, and being more sensitive to the interaction of galaxies with the intergalactic medium and between galaxies.
	
	Two methods are commonly used to detect and measure the HI content of galaxies. Observations of the 21-cm emission line from atomic neutral hydrogen can be used for nearby galaxies. However, the 21-cm emission line is weak at high redshift where HI is more readily detected by using
	ultraviolet absorption against bright background sources. The so-called Damped Lyman-$\alpha$ Absorption (DLA) systems result from absorption by high HI column density (>2$\times$10$^{20}$ atoms cm$^{-2}$) clouds \citep{2005ARA&A..43..861W}.
	
	Targeted 21-cm observations have been carried out on many optically-selected samples \citep{2008ApJ...685L..13C, 1996ApJ...461..609S, 2005ApJ...621..215S}. This method results in efficient measurements of gas in galaxies, but suffers from sample selection limitations, such as the tendency of optical surveys to miss gas-rich low surface brightness galaxies which can have substantial baryon content \citep{1996MNRAS.280..337M, 1997ApJ...482..659D, 1997ApJ...482..104S}. On the other hand, blind HI surveys do not have such selection effects. The drawback is that they require a much larger amount of observing time than optically targeted surveys. Fortunately, fast widefield surveys are now made possible by modern multibeam receivers, phased-array feeds and/or interferometers. For example, \citet{2001MNRAS.322..486B} and \citet{2004MNRAS.350.1195M} used the Parkes 13-beam receiver to conduct a blind survey of the entire southern sky (HIPASS). This was used by \citet{2003AJ....125.2842Z} and others to derive an accurate low-redshift HIMF. \citet{2005AJ....130.2598G} employed ALFA (Arecibo $L$-band Feed Array) with 7 beams to perform a large HI survey in the SDSS region (ALFALFA). \citet{2010ApJ...723.1359M} and others used the data to derive updated HIMF parameters.

	More recent work has investigated the variation of HIMF in different galaxy density environments. \citet{2005MNRAS.359L..30Z} found a steeper faint end slope in higher density regions based on full HIPASS sample \citep{2004MNRAS.350.1195M}. They did not find any clear dependency of the `knee' mass on environment. \citet{2005ApJ...621..215S} used HI spectra of $\sim$ 3000 AGC galaxies to derive a flatter faint end slope and lower characteristic mass in higher density regions. \citet{2014MNRAS.444.3559M} and \citet{2016MNRAS.457.4393J} found a positive correlation between the `knee' mass and density based on the 40\% and 70\% ALFALFA catalogues, respectively. However, neither found convincing evidence for any trend in the low-mass slope. \citet{2018MNRAS.477....2J} found a similar trend in $M^*$ as \citet{2014MNRAS.444.3559M} and \citet{2016MNRAS.457.4393J}, but a steeper $\alpha$ in dense non-cluster regions. \citet{2019MNRAS.486.1796S} found a flatter $\alpha$ in dense regions in the Zone of Avoidance (HIZOA, \cite{2005AJ....129..220D, 2016AJ....151...52S}) survey, but no clear trend in $M^*$. These findings provide information, but as yet no clear overall picture, on the role of interactions, mergers and environment in galaxy formation. Moreover, all the above surveys are limited to the local Universe -- e.g.\ HIPASS ($z<0.04$) \citep{2004MNRAS.350.1195M, 2006MNRAS.371.1855W} and ALFALFA ($z<0.06$) \citep{2005AJ....130.2598G}, so provide no information about the evolution of galaxy properties.

	In this paper, we use the blind HI Arecibo Ultra Deep Survey (AUDS) to focus on galaxies of lower mass and at higher redshift than HIPASS or ALFALFA. The sensitivity of AUDS is 50-200 times better than the above surveys, but is restricted to a much smaller area and cosmic volume. AUDS also uses the Arecibo $L$-band Feed Array (ALFA), but our deeper observation allows us to probe the HIMF up to $z\approx 0.16$. Initial AUDS precursor observations  were used to verify the `drift-and-chase' observational technique, detecting 18 galaxies in redshift range $0.07 < z < 0.16$ \citep{2011ApJ...727...40F}. \citet{2015MNRAS.452.3726H} analysed 60 per cent of the final survey (hereafter AUDS60) to construct a catalogue of 102 galaxies. No significant evolution of the HIMF in the last billion years was found.

	We present the final, complete AUDS survey (AUDS100) and organise the paper as follow. In Section \ref{Observation}, we describe the observations. The data reduction techniques are described in Section \ref{DataReduction}. The source-finding method and galaxy catalogue is described in Section \ref{Source}, where we also investigate completeness and reliability . Two methods for deriving the HIMF are described in Section \ref{Methods}, together with a method to derive corrections for cosmic variance. The HIMF and cosmic HI density $\Omega_{\rm HI}$ derived from the AUDS data is presented in Section \ref{HIMF}. Finally, we discuss and present our conclusions in Sections \ref{Discussion} and \ref{Conclusions}.
	
	In this paper, we use $H_0=70~h_{70}$ km~s$^{-1}$ Mpc$^{-1}$, $\Omega_{\rm M}=0.3$, and $\Omega_\Lambda=0.7$.
	
\section{Observations}\label{Observation}

    In order to avoid solar interference and minimise radio frequency interference (RFI), only night observations were requested. To allow observations during most of the year, we therefore selected two areas located far apart, but still lying within the SDSS footprint \citep{2000AJ....120.1579Y} in order to provide complementary optical information. The field separation also helped to reduce cosmic variance. Regions with strong continuum sources were avoided. The coordinates and effective survey area of the chosen fields are listed in Table \ref{Tab_01}.
    
    Arecibo radio telescope were employed to carry out the observation. The ALFA imporved the speed of the survey by 7 times. The $L$-band beam size of Arecibo telescope is $\sim$ 3.5 arcmin at redshift $z$=0. We used a `drift-and-chase' observation mode as verified by \citet{2011ApJ...727...40F}. The Arecibo telescope was pointed at the western end of the field, a calibration cycle was performed, and the telescope was put in drift mode which allowed the sky to drift through the telescope beams for $\sim$230 s, which corresponds to $\sim1^\circ$ in Right Ascension (RA). Taking into account the ALFA footprint, the coverage in RA is $\sim1.2^\circ$. The procedure was repeated with a shift of 20 arcsec in Declination (Dec). Due to the limited zenith angle range of the Arecibo telescope, a single observation of a field usually took several nights, after which the whole process was repeated.

    Uniform (Nyquist) sampling was achieved by rotating the ALFA receiver by $19.1^\circ$ with respect to the drift direction (see Figure 1 in \citet{2015MNRAS.452.3726H}). In order to maintain bandpass stability, the receiver was not rotated during the observation. Rotation of ALFA was set at the start of the observation according to the observing time and the telescope azimuth and elevation coordinates.
    
    \begin{table}
        \centering
        \caption{The coordinates of two survey fields in AUDS.}
        \begin{tabular}{lccc}
            \hline
            Field    & RA (J2000) range         & Dec (J2000) range    & Area\\
                     & HH:MM:SS                 & DD:MM:SS             & Deg$^2$\\
            \hline
            FIELD17H & 17:02:39.0 to 16:57:21.0 & 19:19:26 to 20:09:26 & 0.68\\
            GAL2577  & 08:22:41.5 to 08:17:18.5 & 21:45:25 to 22:35:25 & 0.68\\
            \hline
        \end{tabular}
        \label{Tab_01}
    \end{table}
    
    The project started in November 2008, and finished in December 2012. Approximately one hour per night was assigned when the target area was overhead. In total, over 1000 hours observation time was assigned for the project, with an on-source time over 700 hours. The remainder was used for calibration or lost due to overheads.
    
    The spectrometer recorded the spectral data in two sub-bands, namely S0 and S1. Each sub-band has a bandwidth of 175 MHz, which is divided into 8192 spectral channels. Each channel corresponds to a frequency resolution of 21.4 kHz and a velocity resolution of 4.5 km~s$^{-1}$ at redshift $z=0$. The two sub-bands were centered at 1.30 GHz and 1.45 GHz, covering a total bandwidth of 300 MHz from 1.225 GHz to 1.525 GHz. These sub-bands overlap a range of 25 MHz. Further observational details can be found in \citet{2015MNRAS.452.3726H}.
    
\section{Data Reduction}\label{DataReduction}

	We used  {\sc livedata} to perform bandpass and flux calibration  \citep{2001MNRAS.322..486B}. It was designed to process multibeam data from HIPASS, but was modified to include the ability to process data from ALFA. The main changes relate to the details of the data format, the Arecibo gain-elevation curve, and the different calibration strategy. HIPASS used a continuous noise-diode calibration, whereas the ALFA calibration cycle produced two short calibration file and one longer data file for each beam and each 230 s scan. 
	
	Due to the high sensitivity of AUDS, RFI was our major problem. The main contributions to RFI at the Arecibo Telescope are Global Navigation Satellite Systems (GNSS) and airport radars. The signals from GNSS have a broad frequency span (many MHz per satellite, mainly from 1.22 to 1.28 GHz) and are continuous in time. Signal strength depends on their position with respect to the horizon and the telescope sidelobe pattern. Radar signals have narrow bandwidth (few MHz) at several frequencies. They are pulsed in time, and not always present. Most RFI is seen in the low-frequency S1 band. The receiver backend/correlator also creates an occasional `birdie' in the central channels of both bands (1.30 GHz and 1.45 GHz). 
	
	Although the {\sc livedata} calibration procedure is robust, and largely immune to RFI and outliers, it only flags strong RFI signals. Therefore \citet{2015MNRAS.452.3726H} developed a flagging script for removing fainter RFI in the AUDS60 data. It performed reasonably well, but an improved flagging technique was required for the higher sensitivity AUDS100 data. Here we detail the updated flagging process:
	
	\begin{enumerate}
	    \item We check the rms noise level in each scan. If the ratio of noise level in the two polarisations is larger than two, we completely remove the data as it is likely to be irrecoverably corrupt due to system problems or very strong RFI.
	    
	    \item We flag data with values deviating by greater than $\pm 0.1$ Jy. After re-calculating the rms, pixels with values exceeding $7\sigma$ are flagged. This procedure removes most RFI generated by the airport radars.
	    
	    \item We remove 2$\sigma$ outliers in the Fourier transform of the temporal values for each frequency channel in each scan. Then we perform an Inverse Fourier Transform. This procedure allows us to remove periodic RFI. 
	    
	    \item We calculate the local rms of scans in the time-frequency plane using three box sizes: $5 \times 5$, $1 \times 5$, and $5 \times 1$ pixels, respectively. If the local rms exceeds three times the global rms in the scan, the pixels will be flagged unless the total flagged percentage for a given spectrum or channel will be less than 80 per cent. Otherwise the spectrum is removed. 
	    
	    \item The ALFA correlator produced a central channel artifact corresponding to 1.3 GHz in band S1. We therefore flagged the central channel in each scan.
	    
	    \item Finally, we checked the flagged fraction in the scan file. If the flagged fraction exceeds 90 per cent, the whole scan is discarded.
	\end{enumerate}
	
	We also remove an artifact caused by {\sc livedata} processing of bright galaxy spectra.  The bandpass calibration procedure, although robust, will slightly depress the scan-average in the frequency range of a bright galaxy, resulting in RA sidelobes. In order to remove this effect, we remove a median value for each channel in each scan file, after masking out any such galaxy profiles.

\begin{figure}
	    \includegraphics[height=5cm,width=8cm]{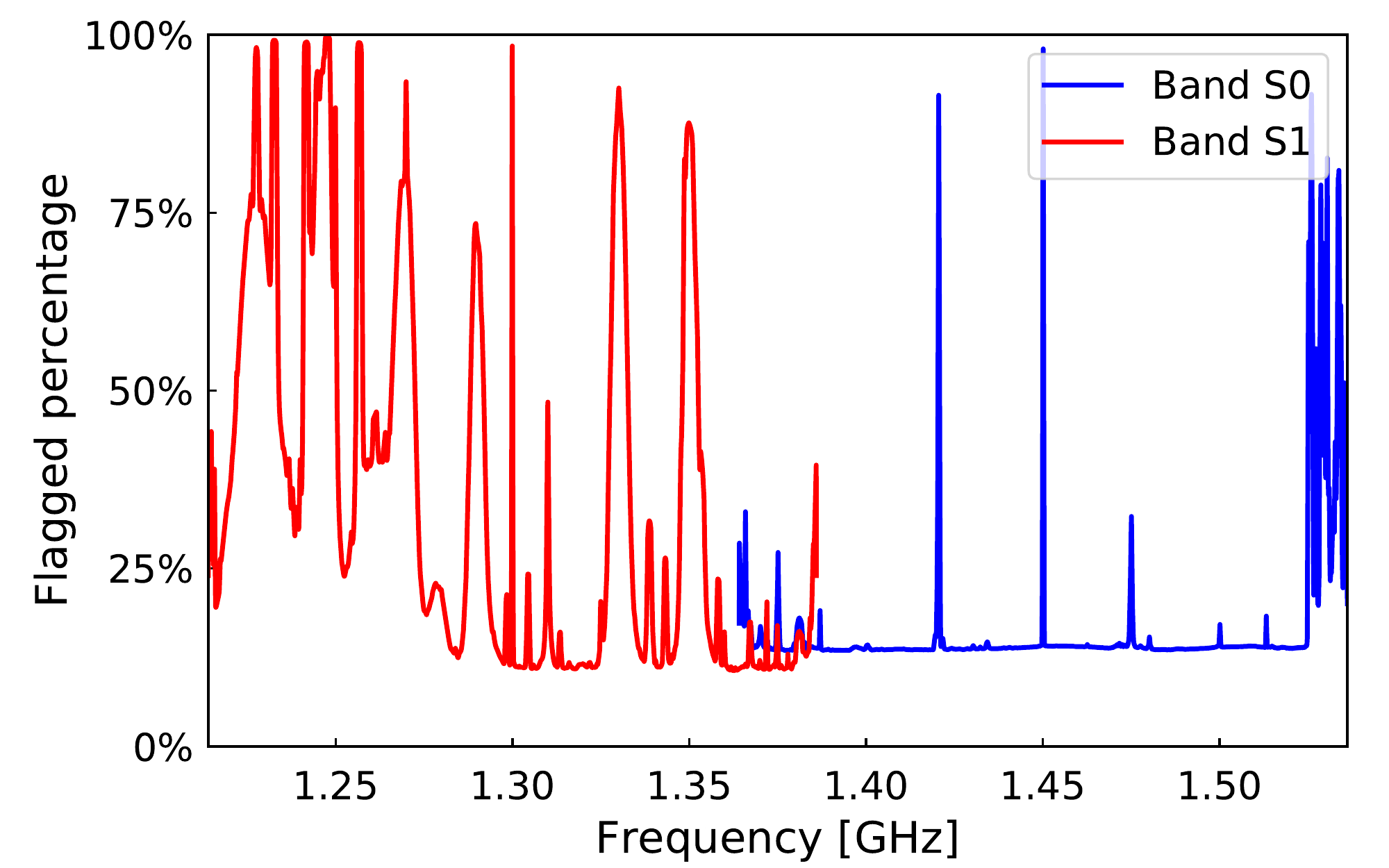}
	    \caption{The flagged percentage in each frequency channel. The blue and red lines represent Band S0 and S1, respectively. RFI mainly resides in Band S1, resulting in a greater percentage of flagged pixels.}
	    \centering\label{Fig_01}
	\end{figure}

	In Figure \ref{Fig_01}, we show the flagged percentage, or RFI occupancy for the two sub-bands. Aside from RFI, the method unfortunately also flags HI emission from the Milky Way, located at 1.42 GHz. However, all the raw data is archived and reusable for future research on HI in Milky Way.
	
	The task {\sc gridzilla} is employed for gridding into image cubes. Since the AUDS100 data volume is too large (8192 channels $\times$ 230 integrations $\times$ 2 polarisations $\times$ 2 sub-bands $\times$ 7 beams $\times$ 11,000 scans) for {\sc gridzilla} to process the whole data set in memory, we divide each spectrum into 64 sub-spectra with an overlap of 22 channels between neighbouring sub-spectra to allow for Doppler shifts. We use {\sc gridzilla} to process these 64 sub-spectra in parallel on a computer cluster. The final cube is generated by linear interpolation of the 64 sub-cubes, resulting in four final cubes, one for each band and each field.

    As each observational period was typically only an hour, relative calibration was based on the receiver noise diode and a template gain-elevation curve for the telescope. The overhead in daily calibrator observations was otherwise too high. Final flux density calibration was therefore adjusted by comparing sources with S/N ratio above 5 in the AUDS area with 46 NVSS continuum sources \citep{1998AJ....115.1693C}. The absolute uncertainty in the final flux density scale using this calibration method was estimated to be 4 per cent. The ratio of integrated fluxes for the only galaxy in common with ALFALFA \citep{2015MNRAS.447.1531C} is $1.02\pm0.03$, in agreement with the above. Note that the flux densities in this paper supersede those in \citet{2015MNRAS.452.3726H} which were on average lower by 29 per cent compared with this paper. Therefore, using the same $h$, AUDS100 on average has a 29 percent higher HI mass than AUDS60.
    
\section{Source Catalogue}\label{Source}

	\subsection{Source-finding methodology}\label{SourceFinding}
	    Automated source-finding software, such as SoFiA \citep{2015MNRAS.448.1922S} is useful in finding both point sources \citep{2017MNRAS.472.4832W} and extended sources (see e.g., \citealp{2015MNRAS.452.2680S, 2019MNRAS.482.3591R}). However, application to AUDS is not straightforward due to the strong variability of the noise level and baseline with position and frequency. We therefore pay close attention to the measurement of the noise level and the flattening of the baseline before passing to  our own custom source-finding algorithm.
	    
	    Due to the dependence on position and frequency, we require a noise estimate for each pixel. If $f_c$ represents the flagged fraction caused by RFI in a given frequency channel in the image cube, then we expect that the noise level for that channel will scale as  $\sigma_c \propto 1/\sqrt{1-f_c}$. Also, if $f_s$ represents the flagged fraction of a given spectrum in the image cube, then the noise level of a spectrum will be expected to scale as $\sigma_s \propto 1/\sqrt{(1-f_s)f_o}$, where $f_o$ is proportional to the observing time at that location. Since a given pixel lies at the intersection of a spectrum and a channel, we might expect the overall noise level at that pixel to be approximated by $\sigma_p \propto 1/\sqrt{(1-f_s)f_o(1-f_c)}$, or $\sigma_p \propto \sigma_s \sigma_c$. This gives a method for deriving an estimate of the noise on the scale of individual pixels. Robust estimates for $\sigma_s$ and  $\sigma_c$ are derived using the consistent estimator $\sigma=1.4826$ MAD, where MAD is the Median Absolute Deviation. 
	    
\begin{figure}
	        \includegraphics[height=7cm,width=8cm]{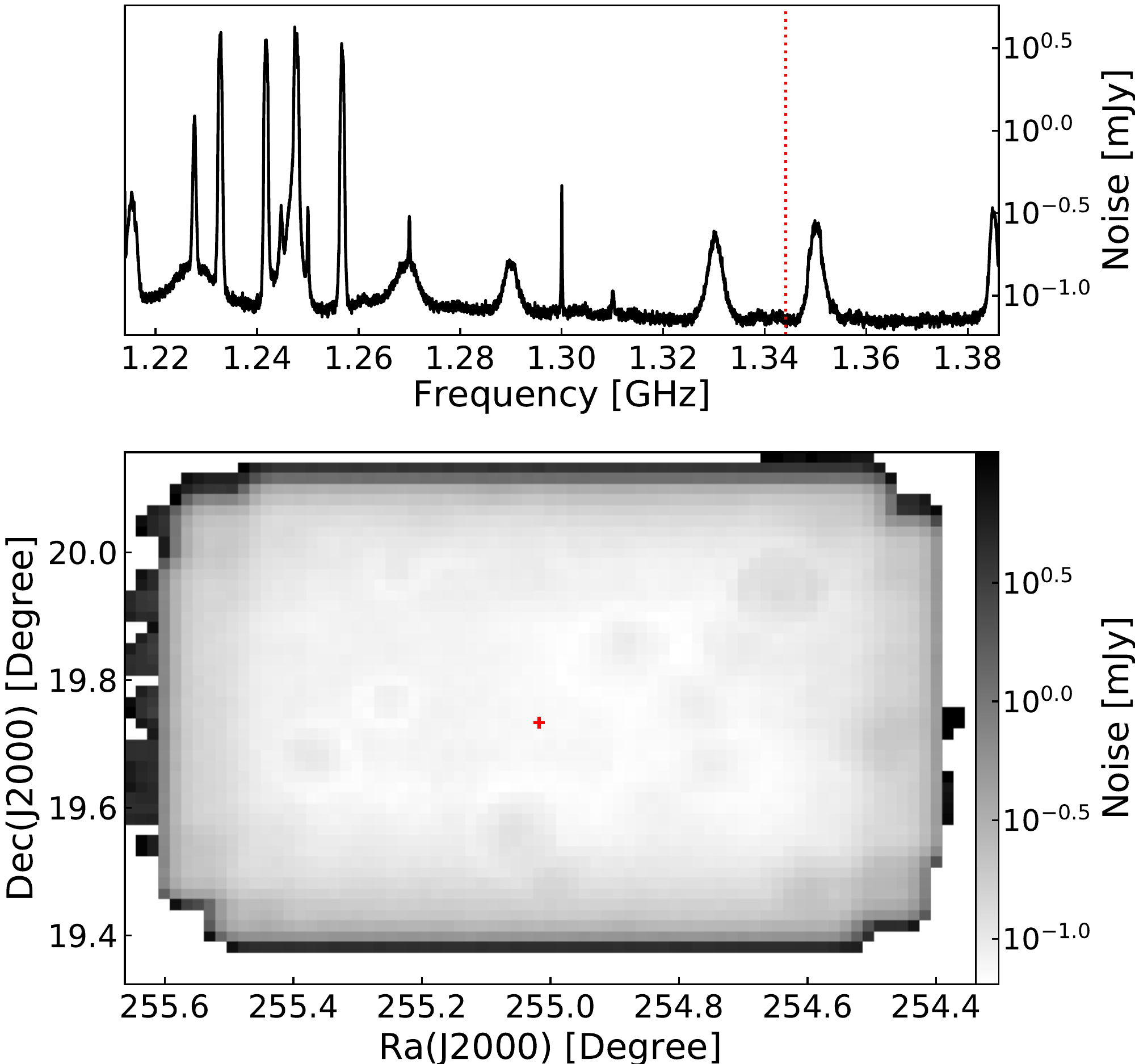}
	        \caption{The estimated pixel noise level $\sigma_p$ as a function of (upper panel) frequency and (lower panel) RA and Dec for FIELD17H Band S1. The rise of $\sigma$ at RFI frequencies is consistent with flagged fraction shown in Figure \ref{Fig_01}. The top panel is a noise spectrum at the field centre RA=255.02$^\circ$, Dec=19.73$^\circ$, which is marked with the red plus in the lower panel. The lower panel is a noise image at 1.34413 GHz, which is indicated by red dotted line in the upper panel. The noise level rises at the edge of the field due to limited observation, and rises due to bad baselines associated with continuum sources.}
	        \centering\label{Fig_02}
	    \end{figure}

	    As an example, we plot $\sigma_p$ as a function of frequency and position for FIELD17H band S1 in Fig.~\ref{Fig_02}. It can be seen that $\sigma_p$ rises at highly-flagged frequencies between 1.22 and 1.28 GHz. It also rises at the edge of the field due to limited observational coverage, and rises due to bad baselines associated with continuum sources. 
	    
        For baseline fitting, we firstly smooth each spectrum in the cube in successive steps up to 20 channels (0.43 MHz). For each smoothing length, we fit increasingly high-order polynomial baselines, starting at 0 and ending at 24 (over the whole 175 MHz). At each step, we re-calculate the noise level for the smoothed cube, record points deviating by more than $3\sigma$ as candidate sources, and remove those points from the subsequent baseline fitting procedure. Spectra with bad baselines may end up being 100 per cent flagged in this procedure. RFI will only cause a limited number of channels to be flagged. Candidate sources extending less than 3\arcmin\ spatially and 128 kHz (27 km~s$^{-1}$) spectrally are removed from the list. 
	    
	    All sources were checked by eye and several obviously false detections were removed, resulting in 153 candidates. We then inspected the data cube by eye for additional candidates. The criteria were either: 1) the candidate has the width of a beam and is detected in one or multiple channels when viewed in the position-position plane; or 2) the candidate is detected in a 3$\times$3 pixel spatially-averaged spectrum. This process picks up a further 94 candidates with S/N ratios slightly less than 3$\sigma$. These candidates are of lower reliability (see Section~\ref{Reliability}) hence it is likely that some are false detections. Due to the large beamwidth ($\sim$ 3.4 arcmin) of Arecibo at this frequency, some sources are confused. We inspect the spectral and spatial morphology of these sources, and attempt to separate them into individual sources.
	    
	\subsection{Catalogue}
	
\begin{figure*}
	        \includegraphics[height=9cm,width=18cm]{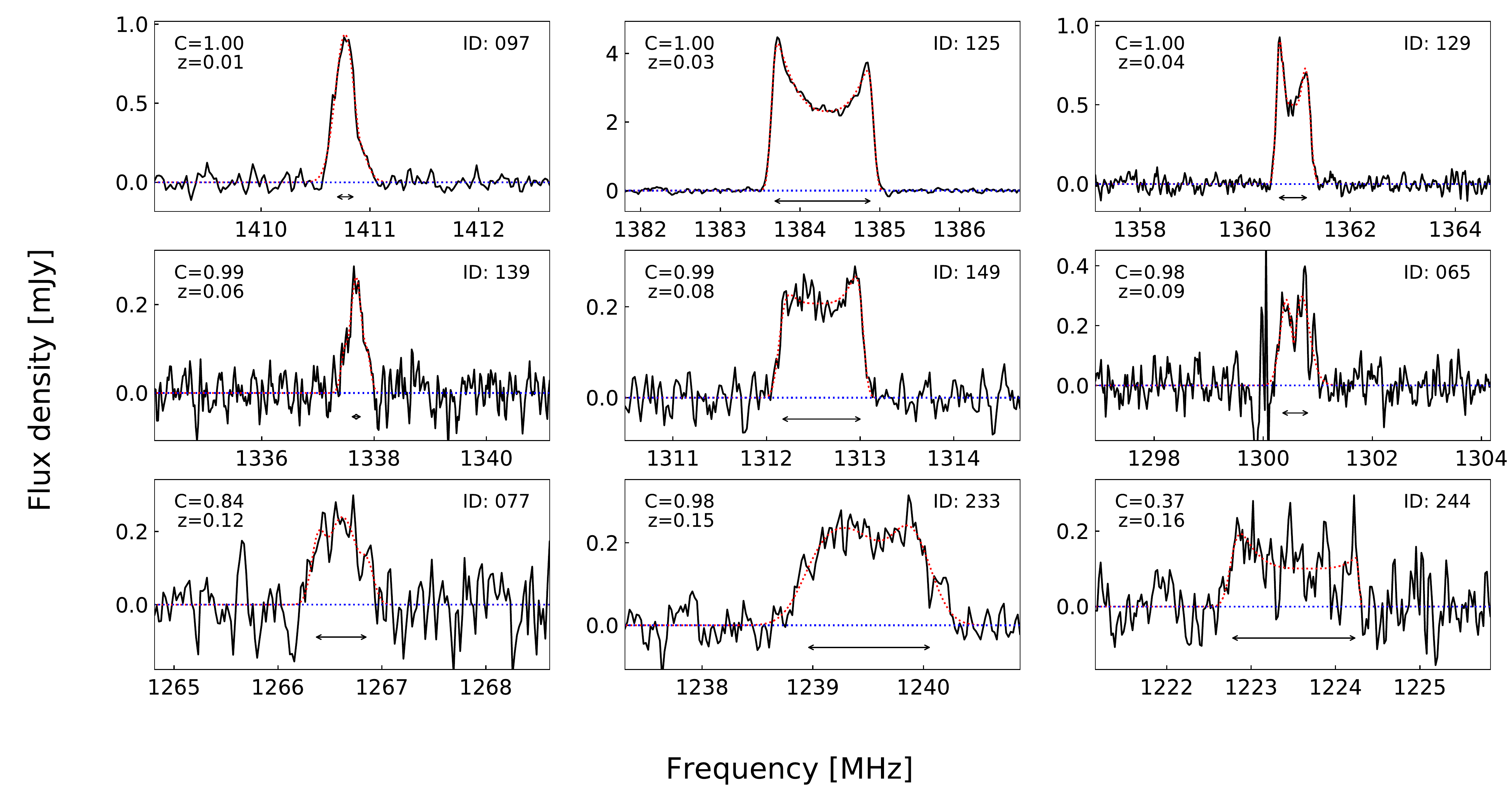}
	        \caption{Example spectra of nine AUDS100 galaxies at different redshifts. The red dotted lines show the best-fit Busy functions; the blue dotted lines indicate zero flux density; the double-sided arrows indicate the $W_{50}$ linewidth. The completeness parameter $C$, defined in Section~\ref{Completeness}, and redshift is listed for all detections. The spectra for all galaxies are available online.}
	        \centering\label{Fig_03}
    	\end{figure*}

	    We detect 247 galaxies in total, of which 9 are detected in both S0 and S1 bands. We use AUDS100 to indicate all AUDS galaxies, and AUDS100$_{\rm Pro}$ to indicate 153 AUDS galaxies only detected by our own custom source-finding algorithm. The 94 AUDS galaxies found by eye are indicated by AUDS$_{\rm Eye}$. For each source, we use a cubic polynomial to fit the baseline. A two-dimensional Gaussian is used to fit the moment map in order to derive the source central position. We assume that the galaxies we detect are not resolved. The mean and median FWHM of fitted two-dimensional Gaussian function is 3.42 and 3.46 arcmin, respectively, which is consistent with this assumption. Spectra within the FWHM ellipse are used to determine the spatially integrated spectrum for each galaxy:
	
	    \begin{equation}\label{Equ_01}
	    	S_i(\nu)=\frac{\Sigma_{xy} s_i(x, y, \nu)}{\Sigma_{xy} f(x-x_i,y-y_i) } ,
	    \end{equation}
	    where $s_i(x,y,\nu)$ is the flux density of the voxel at $(x, y, \nu)$ of galaxy $i$, and $f(x,y)$ is the 2 dimensional Gaussian function fitted to galaxy $i$ with amplitude of unity, and centred on the fitted galaxy position $(x_i, y_i)$. A fifth-order polynomial is also fitted and removed from the spatially integrated spectrum $S_i(\nu)$. For the 9 galaxies detected in both bands, the spectrum with lowest noise level is used in our final result. Example spectra over a range of redshifts are shown in Figure \ref{Fig_03}.
	    
	    The flux-density-weighted frequency is used to derive the redshift of each galaxy. In order to remove the influence of noise on properties of galaxy, we use the Busy Function \citep{2014MNRAS.438.1176W} to fit the spatially integrated spectra, then obtain the frequency-integrated flux $S_{\rm int}$, the peak flux density $S_{\rm peak}$, and linewidth $W_{50}$. The HI mass is defined using:

	    \begin{equation}\label{Equ_02}
	    	M_{\rm HI}~[h_{70}^{-2}{\rm M_\odot}] \simeq 49.7 \times D_{\rm L}~[h_{70}^{-1}{\rm Mpc]}^2 S_{\rm int}~[\rm Jy~Hz]
    	\end{equation}
        \citep{2017PASA...34...52M}, where $D_{\rm L}$ is the luminosity distance. We list the properties of the galaxies in Table \ref{Tab_02}. 
    	
        The common detected sources in both bands allow us to estimate a lower bound on the uncertainties in coordinates and recession velocity. The mean spatial and recession velocity differences are $0.22\pm 0.20$ arcmin and $11\pm 17$ km~s$^{-1}$, respectively. Considering the spatial resolution of $\sim$3.4 arcmin and velocity resolution $\sim$4.5 km~s$^{-1}$, these uncertainties are acceptable.
    	
    	We compare the AUDS100 catalogue with the AUDS60 catalogue \citep{2015MNRAS.452.3726H}. Using a maximum spatial and velocity separation of 3 arcmin and 250 km~s$^{-1}$, respectively, 86 out of 102 AUDS60 sources have a counterpart in AUDS100. A further two sources (J081933+220624, J082011+223025) were found to have separations of more than 3 arcmin. Two confused pairs of AUDS60 sources (J165831+200304 and J165832+195702, J170122+195245 and J170123+195238) appear to correspond to only two AUDS100 sources. One source (J165659+201232) is outside the nominal survey boundary. Two weak sources (J165952+193840 and J170117+200235) have very narrow line widths, and are probably RFI artifacts. Another AUDS60 source (J170133+200336) also lies in a region with strong RFI. The remaining 8 AUDS60 sources (J081749+221114, J081957+221628, J082034+220848, J082116+221604, J082130+221945, J165938+193030, J165938+193512, J170133+193253) are not detected in AUDS100 and likely not real. 
        \vspace*{\fill}
        \begin{sidewaystable}
            \small
            \begin{minipage}{240mm}
                \caption{\quad Properties of galaxies in AUDS100. The full version can be found online.}
                \begin{tabular}{ccccccccccccc}
                    \hline
                    AUDS ID & RA, Dec (J2000)            & $cz$            & $S_{\rm int}$ & $S_{\rm peak}$  & $W_{50}$ & $\log(M_{\rm HI}h_{70}^2/{\rm M}_\odot)$         & C    & Sample  & Pro & G & AUDS60                         &  RA Dec (J2000) (OC)      \\
                            & HH:MM:SS $\pm$ DD:MM:SS & km~s$^{-1}$   & mJy~MHz    & mJy       & MHz    &  &      &         &     &   &                                & (HH:MM:SS $\pm$ DD:MM:SS) \\
                            (1) & (2) & (3) & (4) & (5) & (6) & (7) & (8) & (9) & (10) & (11) & (12) & (13) \\ 
                    \hline
					001 & 16:58:22.3+19:55:13     & 1838.6  & 0.094  & 0.42  & 0.23 & 6.51  & 0.95 & ...   & N & -  & ...                            & ...                     \\
					002 & 16:59:11.8+20:06:15     & 4346.9  & 0.187  & 1.50  & 0.12 & 7.56  & 0.96 & ...   & N & -  & ...                            & ...                     \\
					003 & 16:58:02.2+19:54:07     & 4389.1  & 0.220  & 0.85  & 0.26 & 7.64  & 1.00 & $Sub$ & Y & -  & J165806+195322                 & 16:58:02.4+19:53:60$^2$ \\
					004 & 16:58:39.2+19:56:11     & 4682.9  & 0.291  & 0.30  & 1.04 & 7.82  & 0.88 & ...   & N & -  & ...                            & ...                     \\
					005 & 16:58:41.3+19:34:12     & 5631.2  & 0.082  & 0.34  & 0.22 & 7.43  & 0.96 & $Sub$ & Y & -  & ...                            & ...                     \\
					006 & 16:59:34.2+19:45:21     & 6012.7  & 0.608  & 1.84  & 0.32 & 8.36  & 1.00 & $Sub$ & Y & -  & J165934+194525                 & 16:59:33.6+19:45:36$^2$ \\
					007 & 16:58:29.8+20:02:46     & 6022.5  & 12.985 & 27.79 & 0.47 & 9.69  & 1.00 & $Sub$ & Y & -  & J165831+200304, J165831+200304 & 16:58:29.6+20:02:29$^1$ \\
					008 & 17:02:13.7+19:29:56$^b$ & 9417.6  & 1.915  & 5.73  & 0.33 & 9.26  & 1.00 & $Sub$ & Y & -  & J170212+193010                 & ...                     \\
					009 & 16:59:51.3+19:44:28$^b$ & 9774.7  & 0.322  & 0.77  & 0.44 & 8.52  & 1.00 & $Sub$ & Y & -  & J165950+194442                 & 16:59:52.8+19:44:24$^2$ \\
					010 & 17:00:18.4+19:31:39$^b$ & 9805.4  & 0.176  & 0.39  & 0.43 & 8.26  & 0.97 & $Sub$ & Y & -  & J170015+193342                 & 17:00:19.2+19:31:48$^2$ \\
                    \hline
                \end{tabular}
                \label{Tab_02}
                Cols (1)-(2): AUDS100 identification, RA and Dec (J2000) coordinates from the two dimensional Gaussian fit, respectively. Col (3): Recession velocity, $cz$. Cols (4)-(6): $S_{\rm int}$, $S_{\rm peak}$, and $W_{50}$ - integrated flux, peak flux density and linewidth, respectively. Col (7): HI mass. Col (8): completeness estimation of each galaxy (see Section~\ref{Completeness}). Col (9): sample designation. Col (10): whether source is detected by the custom source finding algorithm. Col (11): Flag for confusion. Col (12): identification of AUDS60. Col (13): coordinates of the optical counterpart (see \citealp{2014UWAThesisH}).\\
                $^b$: Galaxies detected in both band S0 and S1.\\
                $^1$: SDSS \\ 
                $^2$: AUDSOC \citep{2014UWAThesisH}.\\
            \end{minipage}
        \end{sidewaystable}

    \subsection{Optical counterparts}
    \label{s:optical}
        
        The two fields of AUDS are covered by the SDSS. By cross-matching with the SDSS DR14 \citep{2018ApJS..235...42A}, we find that there are 191 optical galaxies with spectroscopic redshifts  in FIELD17H, and 598 in GAL2577. \citet{2014UWAThesisH} did a follow up spectroscopic observation of an additional 787 galaxies using the Anglo-Australian Telescope (AAT). These galaxies were selected from the SDSS DR7 catalogue based on $r$ < 20.3 mag in the AUDS survey area. Of these, there are 229 spectra in the AUDS redshift range of sufficient quality.
        
\begin{figure}
	        \includegraphics[height=5cm,width=8cm]{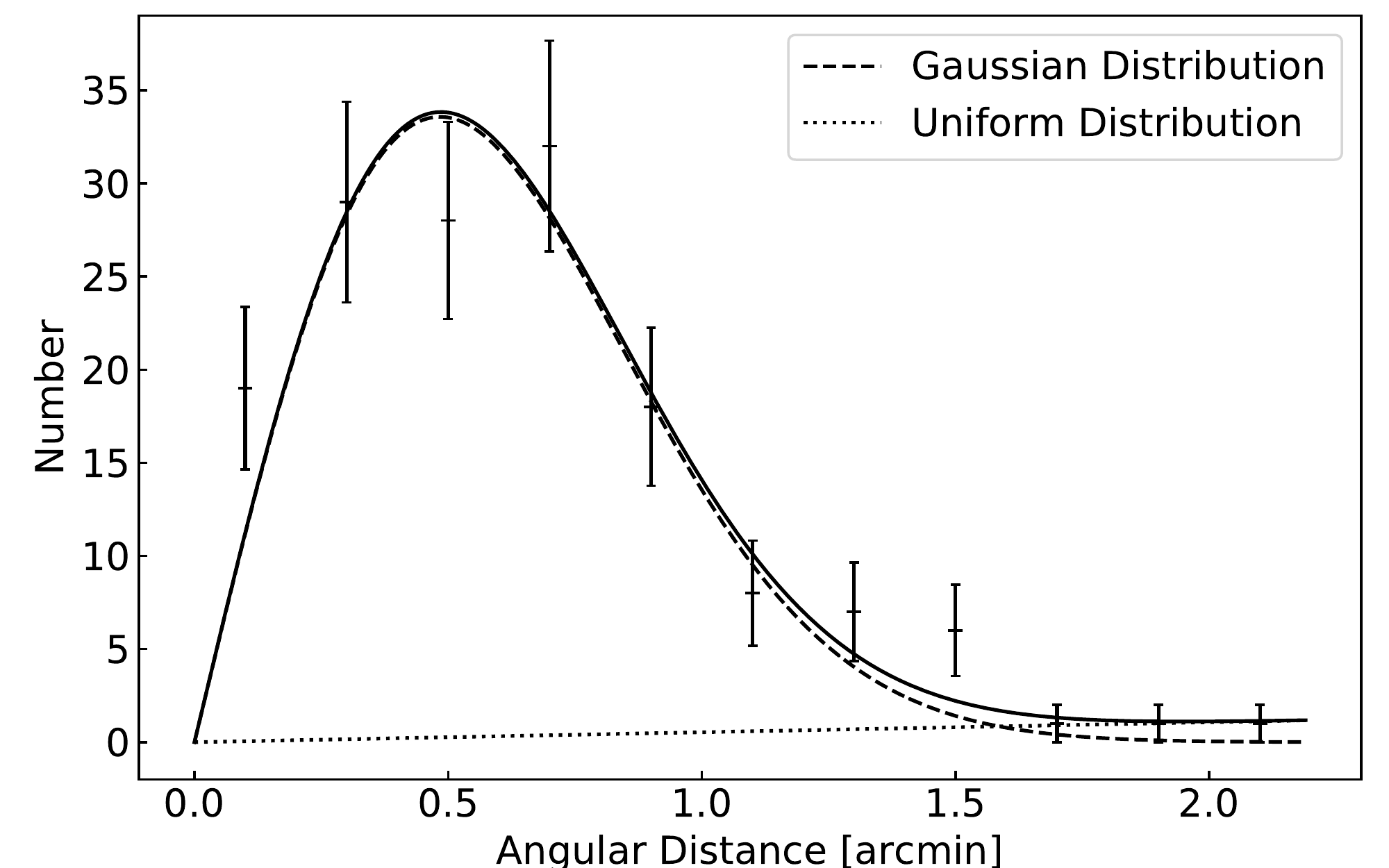}
	        \caption{The distribution of angular distances between the HI and optical positions of the 150 AUDS100 galaxies without optical spectroscopic redshifts. The AUDS100 galaxies are matched to the nearest SDSS galaxy with a photometric redshift $z<0.3$.  Poisson error bars are shown. The dashed line represents genuine matches with a 2D Gaussian distribution of position errors, while the dotted line represents a confusing component with a uniform distribution. The solid line is the summation of the two components.}
	        \centering\label{Fig_04}
    	\end{figure}

        Optical counterparts are matched to AUDS100 detections when the angular distance is within 1.8 arcmin, and the velocity difference is within 200 km~s$^{-1}$. If an optical galaxy matches more than one HI source, the nearest one is chosen as the counterpart. In total, 97 out of the 247 galaxies have optical counterparts. For the rest of the 150 galaxies, we search SDSS for galaxies in the AUDS area with photometric redshifts less than 0.3 and angular separation within 5 arcmin. With these criteria, we find optical counterparts for all the remaining 150 galaxies.  The distribution of angular distance is shown in Figure \ref{Fig_04}. All photometric counterparts lie within 2.1 arcmin, indicating that most of them are likely to be genuine associations. We fit a two-component function consisting of a Gaussian distribution and an uniform distribution, both in angular coordinates. The Gaussian distribution represents the genuine associations, while the uniform distribution represents confusing sources at different redshift. The result indicates that there may only be $\sim$ three sources (AUDS ID: 112, 186, 206) whose optical counterparts or HI detections are in doubt. Other matches all lie within 1.6 arcmin of angular distance. Further optical observations are required to confirm these associations.
        
    \subsection{Completeness}\label{Completeness}
        
	    \begin{figure}
	        \includegraphics[height=5cm,width=8cm]{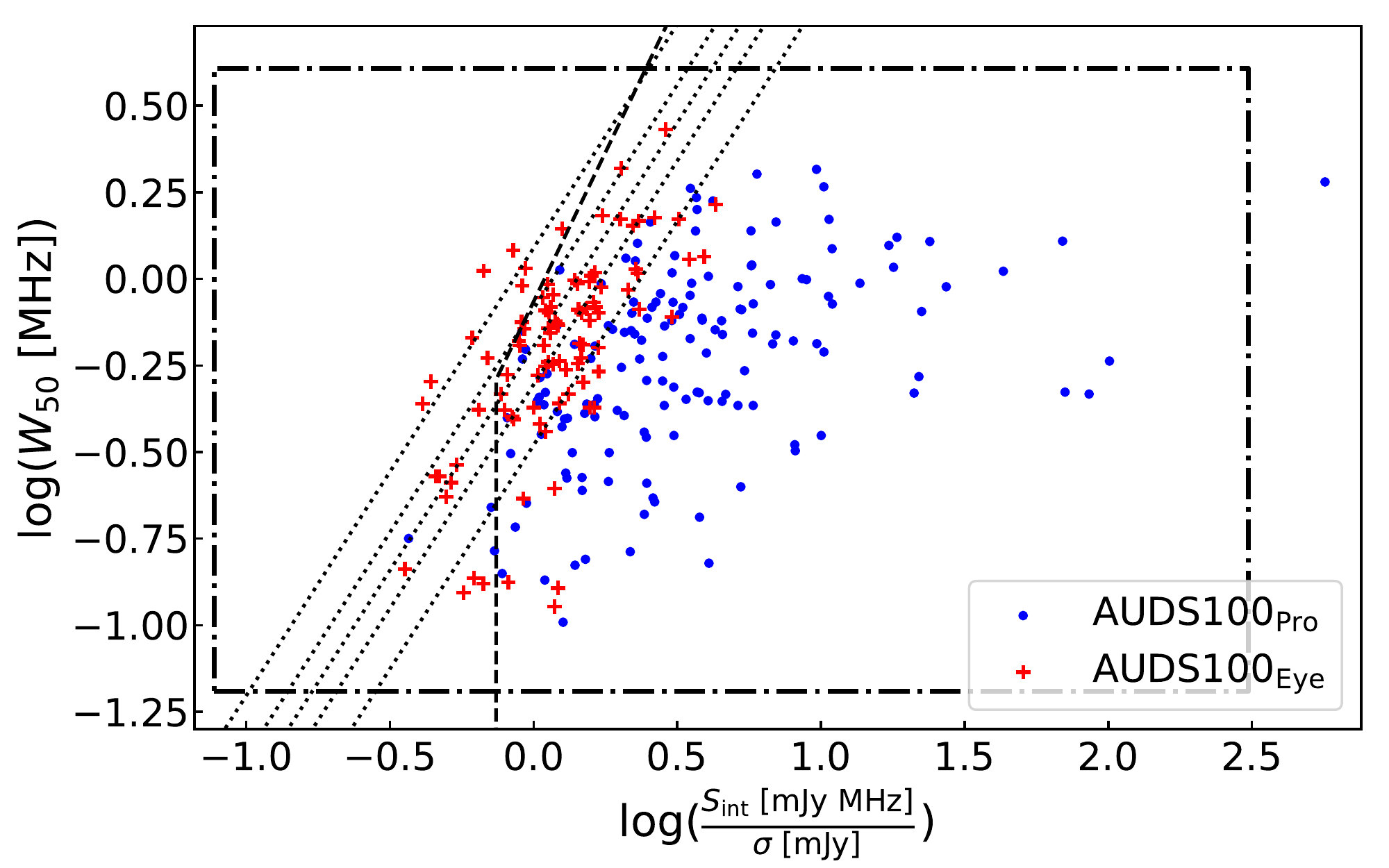}
	        \caption{The distribution of AUDS100 galaxies in the $\log(S_{\rm int}/\sigma)$-$\log(W_{50})$ plane. The blue dots represent galaxies automatically detected (by our algorithm), while the red plus signs represents galaxies identified by eye. The dot-dashed box shows the area of parameter space used in our completeness simulation. The dotted lines represent completeness levels of 0.1, 0.3, 0.5, 0.7, and 0.9 (from left to right) as determined from the simulation, while the dashed line represents the completeness limit derived from the $T_C$ method (see Section \ref{TcMethod}).}
	        \centering\label{Fig_05}
    	\end{figure}
        
\begin{figure}
	        \includegraphics[height=5cm,width=8cm]{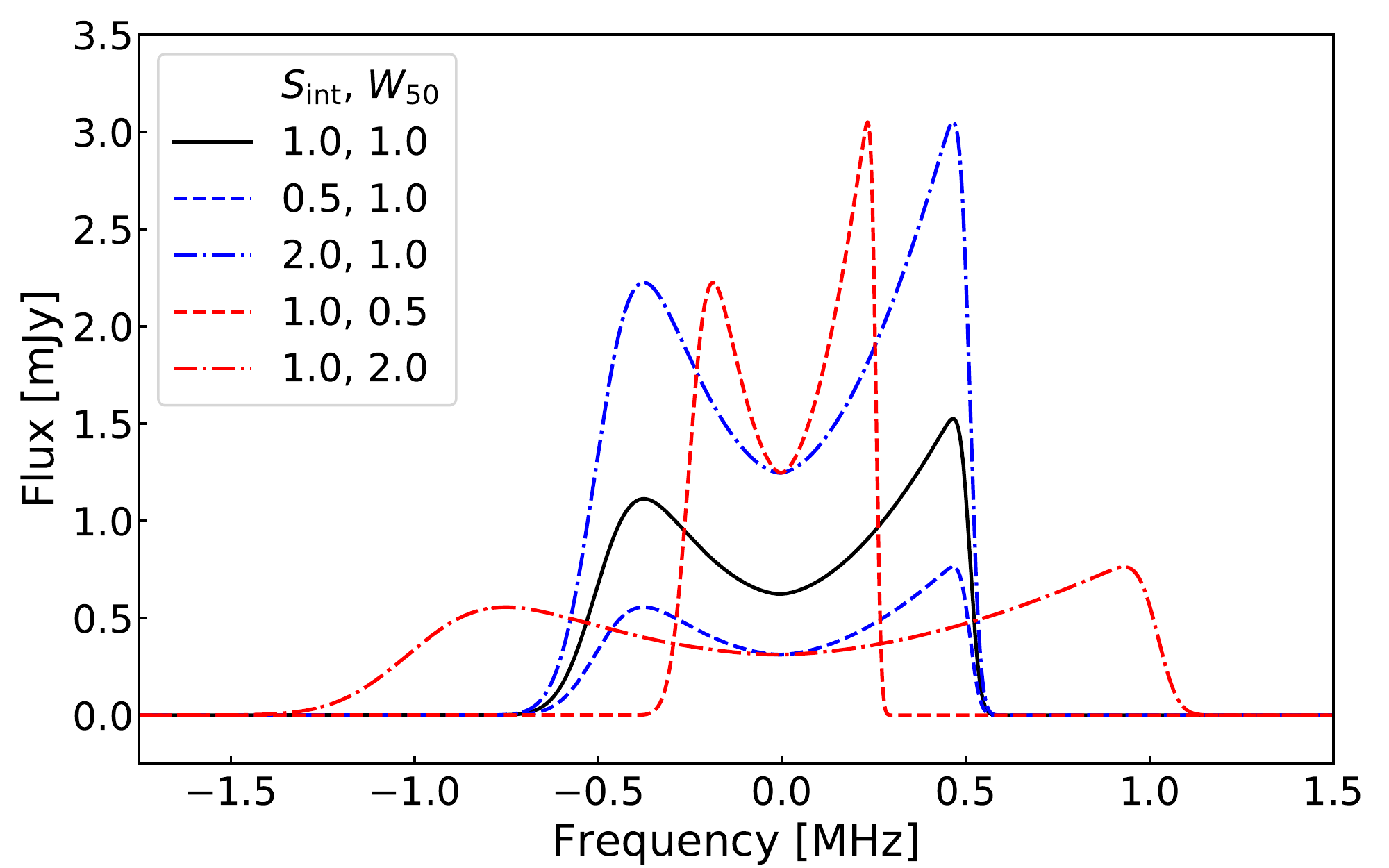}
	        \caption{An example spectral profile of an AUDS100 galaxy is shown by the black solid line. The red profiles have the same integrated flux, but are scaled to a different linewidth. The blue profiles have the same linewidth, but are scaled to different fluxes. Profiles such as these are used as templates for creating synthetic galaxies in order to measure sample completeness. In the legend, $S_{\rm int}$ and $W_{50}$ are in the units of mJy~MHz and MHz, respectively.}
	        \centering\label{Fig_06}
    	\end{figure}

        Completeness can be derived by inserting synthetic galaxies into the data cubes as previously described by \citet{2004MNRAS.350.1210Z} and \citet{2015MNRAS.452.3726H}. The latter authors found that the AUDS completeness could be well characterised in the $\log(S_{\rm int}/\sigma) - \log(W_{50})$ plane, rather than the $\log(S_{\rm int}) - \log(W_{50})$ plane. We follow the same procedure. The synthetic galaxies are generated with a range of noise-scaled integrated fluxes ($-1.11 < \log(S_{\rm int}{\rm [mJy~MHz]}/\sigma{\rm [mJy]}) < 2.49$) and linewidth ($-1.19 < \log(W_{50}{\rm [MHz]}) < 0.61$), as shown in Figure \ref{Fig_05}. The completeness is related to the shape of line profile as well as $(S_{\rm int}/\sigma)$ and $W_{50}$. We therefore randomly select our Busy function fits to the AUDS100 profiles, using $S_{\rm int}$ and $W_{50}$ to modify the profile appropriately. Figure \ref{Fig_06} shows an example. We insert 15 synthetic galaxies into each sub-frequency interval (10 MHz) from 1420 MHz downward, using random positions and frequencies without overlapping with known galaxies in the data cubes. The noise level $\sigma$ for each synthetic galaxy was obtained from the data cube once the position and frequency were known. A Gaussian function of FWHM 3.4 arcmin was convolved with the model.
        
\begin{figure}
	        \includegraphics[width=8.0cm, height=6.0cm]{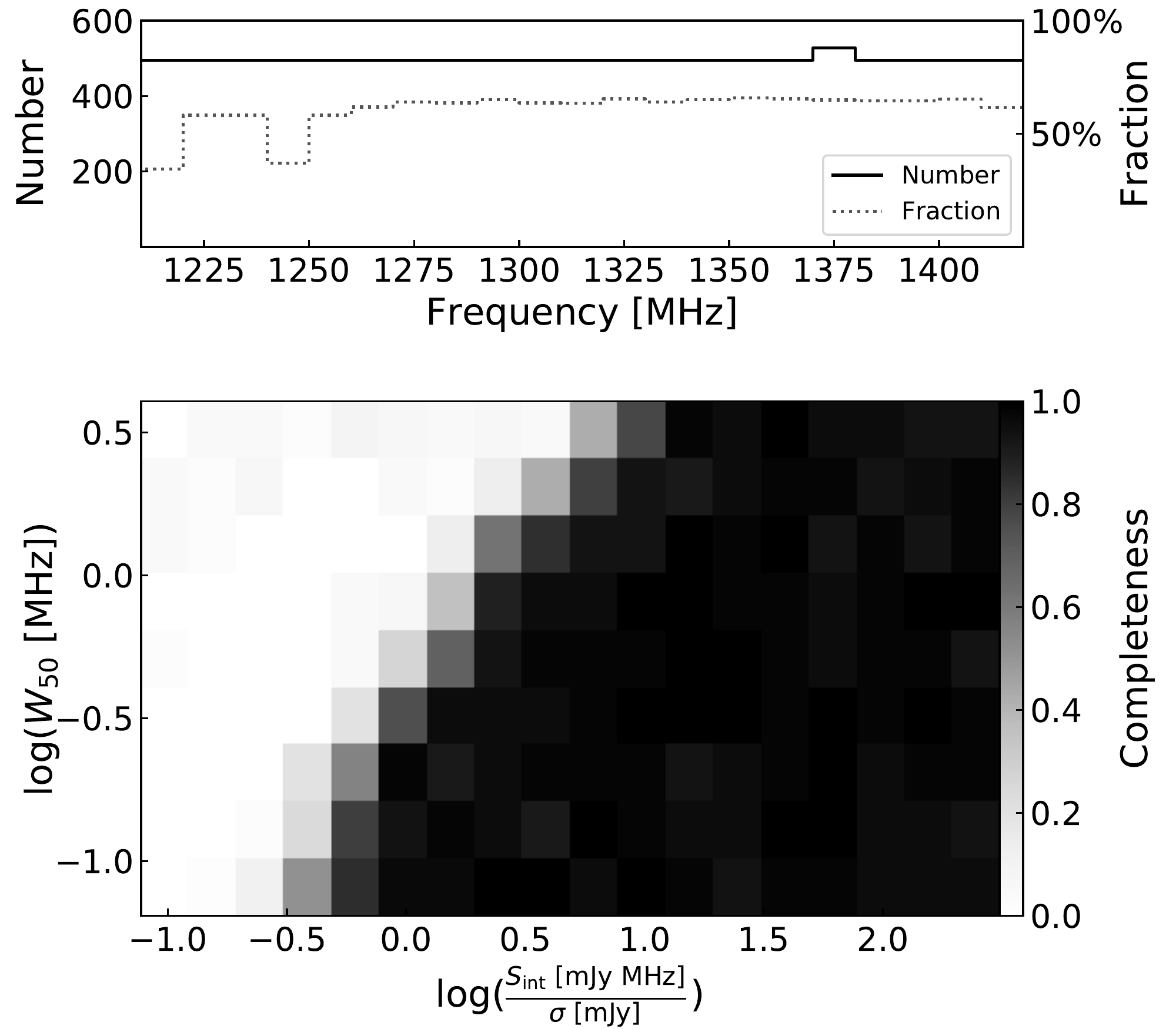}
	        \caption{Top panel: the number of synthetic galaxy profiles inserted into the data cube at each frequency bin, and the detection rate of our automated source-finder. Bottom panel: the detection rate in each cell, where black is 100 per cent and white is 0 per cent.}
	        \centering\label{Fig_07}
    	\end{figure}

        We use the automated procedure of Section \ref{SourceFinding} to search synthetic galaxies. This can give us the completeness for AUDS100$_{\rm Pro}$ sample. The number of synthetic galaxies in each frequency bin, and the detection rate is shown in Figure \ref{Fig_07} (top panel). Since there is an overlap between the S0 and S1 bands around 1375 MHz, the synthetic galaxy number is a little higher at that frequency. The detection rate is reasonably constant at $\sim$ 60 per cent in the different frequency bins. However, it drops dramatically at 1215 and 1245 MHz, where there is strong RFI. The detection rate for each cell in the $\log(S_{\rm int}/\sigma) - \log(W_{50})$ plane is also shown in Figure \ref{Fig_07} (bottom panel). It shows clearly that the detection rate is 0 per cent for weak galaxies with broad linewidth, and 100 per cent for strong galaxies with narrow linewidth.
        
\begin{figure}
	        \includegraphics[height=6cm,width=8cm]{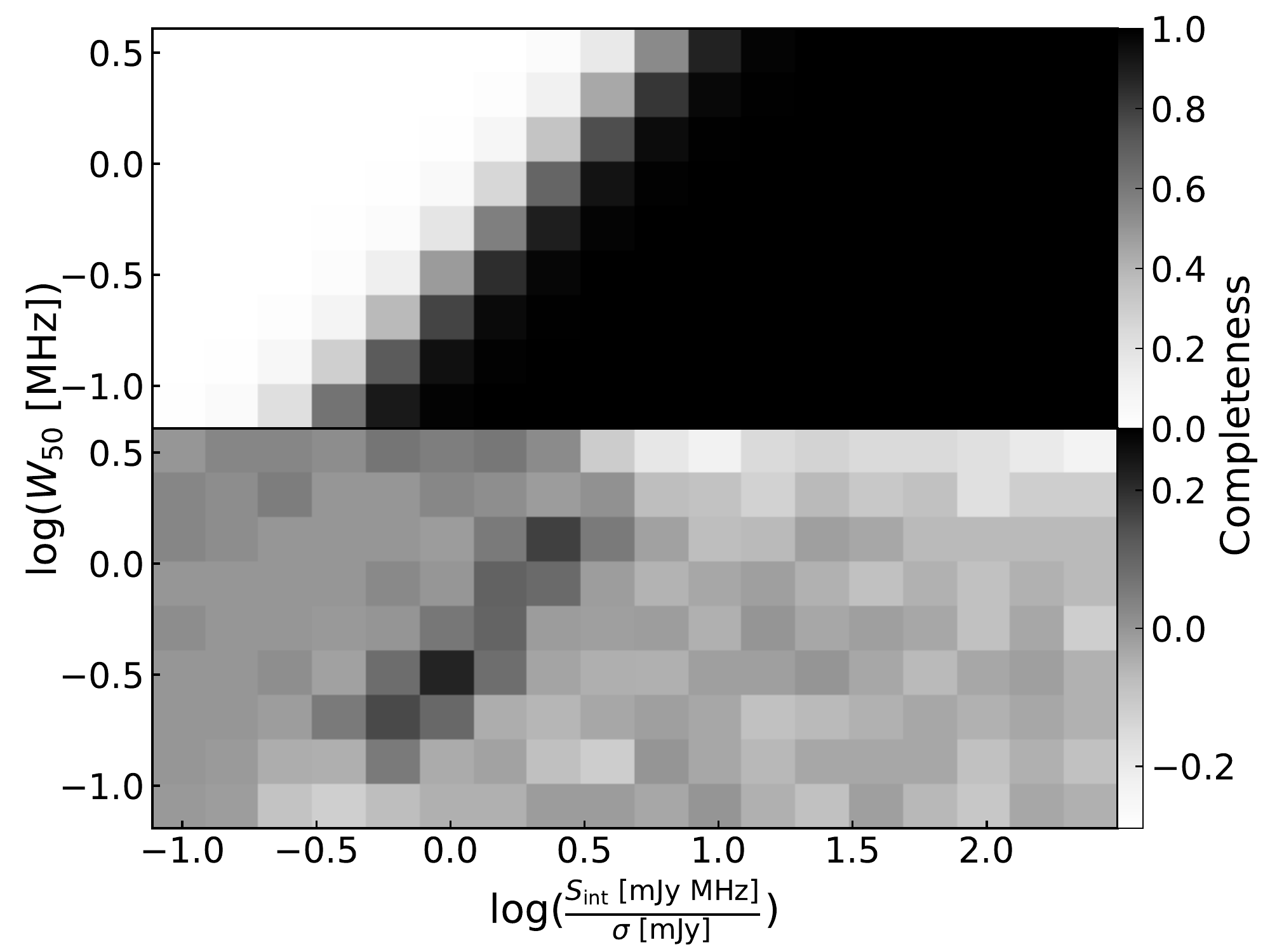}
	        \caption{The two dimensional sigmoid fit of Equation~\ref{eq:sigmoid}, using parameters from Table~\ref{Tab_03}. The upper panel shows the mean value in each cell from the completeness model, while the lower panel shows the residual.
	        }
	        \centering\label{Fig_08}
    	\end{figure}

    	\citet{2015MNRAS.452.3726H} used the error function to model the completeness function. In this paper, we use a simple sigmoid function:
    	
    	\begin{equation}
    	\label{eq:sigmoid}
    	    C=\left(1+\exp{\left(-C_0 \log\left(\frac{S_{\rm int}}{\sigma}\right)+C_1 \log(W_{50})+C_2 \right)}\right)^{-1}
    	\end{equation}
    	where $S_{\rm int}$ is the integrated flux in mJy~MHz, $\sigma$ is the local rms in mJy, $W_{50}$ is the linewidth in MHz, and $C_0$, $C_1$, and $C_2$ are the free parameters of the fit. After accounting for Poisson errors in each bin, the fitted results are listed in Table \ref{Tab_03}. The model and residual is shown in Figure \ref{Fig_08}. There are some minor differences ($\pm 0.3$) near the steepest part of the completeness function. The same procedure is performed for the higher-reliability {\it Sub$_{\rm Pro}$} sample (defined in Section \ref{Reliability}). 
    	
    	\begin{table}
    	    \centering
    	    \caption{The fit parameters for completeness models describing the full AUDS100 sample and the high-reliability {\it Sub} samples. Subscript `Pro' refers to the sub-sample of galaxies detected by our source-finding program.}
    	    \begin{tabular}{lccc}
    	        \hline
    	        Sample & $C_0$ & $C_1$ & $C_2$ \\
    	        \hline
    	        AUDS100$_{\rm Pro}$ & 9.98(0.73)  & 7.72(0.59) & 3.76(0.22)  \\
    	        AUDS100             & 9.98(0.73)  & 7.72(0.59) & 1.51(0.12)  \\
    	        $Sub_{\rm Pro}$     & 9.95(0.83)  & 7.66(0.67) & 3.74(0.25)  \\
    	        $Sub$               & 9.95(0.83)  & 7.66(0.67) & 3.72(0.12)  \\
    	        \hline
    	    \end{tabular}
    	    \label{Tab_03}
    	\end{table}
    	
    	As we mentioned, the result can only give us the completeness for the AUDS100$_{\rm Pro}$ sample found by our source-finding program, and the high-reliability {\it Sub}$_{\rm Pro}$ sample. However, in order to estimate completeness for the full AUDS100 sample, which includes by-eye detections, we made an assumption that the completeness is also described by Equation \ref{eq:sigmoid}, but with a shift described by a modified $C_2$ parameter. 
    	
\begin{figure}
	        \includegraphics[width=8cm,height=5cm]{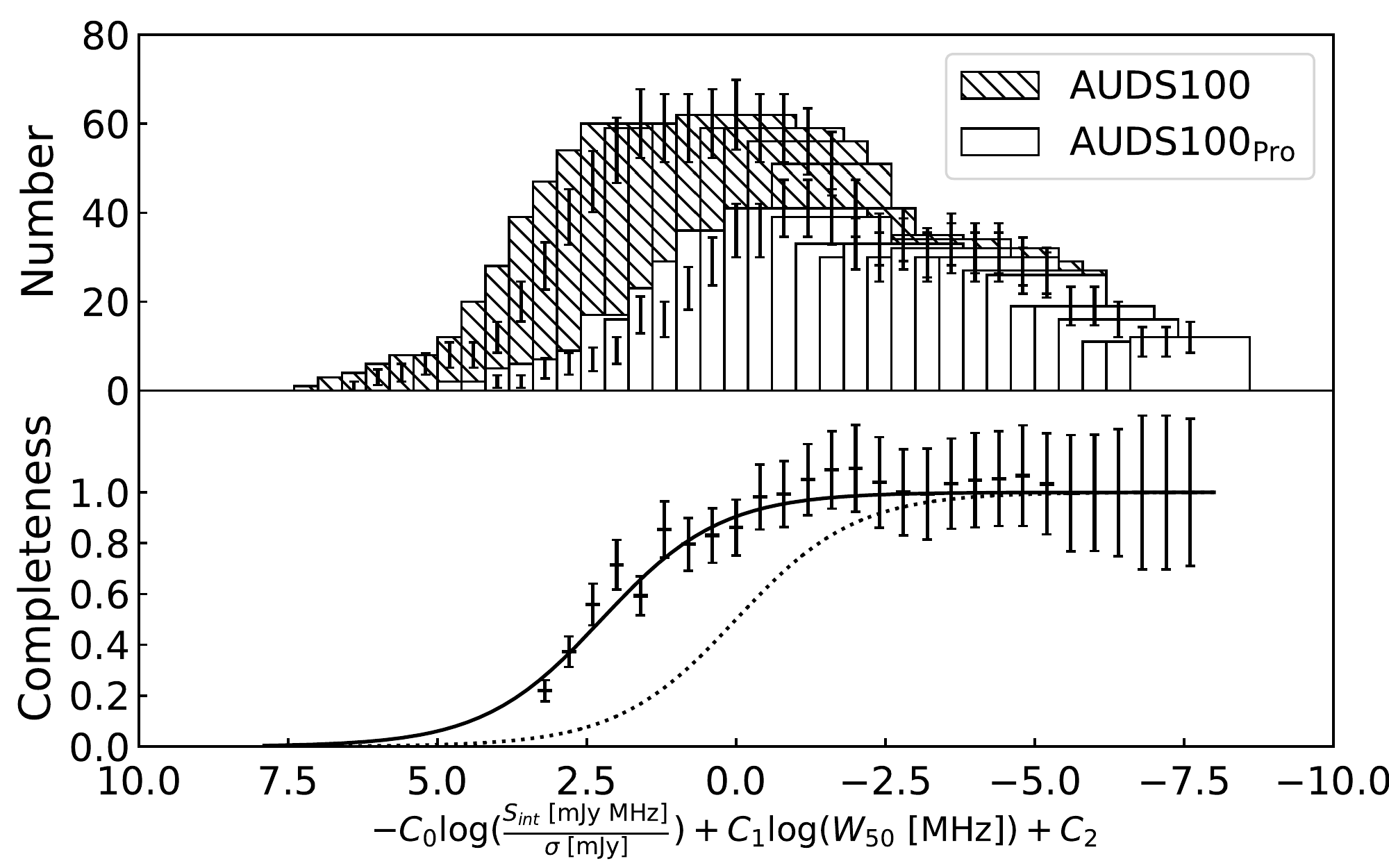}
	        \caption{In the upper panel, the hatched and open histograms represent galaxy numbers in each bin in the AUDS100 and AUDS100$_{\rm Pro}$ samples, respectively. The bin width and spacing are 2.0 dex and 0.4 dex, respectively. In the lower panel, the completeness for AUDS100$_{\rm Pro}$ and AUDS100 is shown by the dotted and solid lines, respectively.}
	        \centering\label{Fig_09}
    	\end{figure}

    	We then re-bin the data at different values of $-C_0 \log(S_{\rm int}/\sigma)+C_1 \log(W_{50})+C_2$ such that the galaxies in the same bin have same completeness for one sample.. The total number of galaxies that should be in each bin, corrected for completeness, is then $N_{\rm Bin}/C_{\rm Bin}$, where $N_{\rm Bin}$ is the galaxy number in that bin, and $C_{\rm Bin}$ is the completeness. The total number in each bin should be the same, no matter in which sample:
    	
    	\begin{equation}
    	\frac{N_{\rm Bin}[{\rm AUDS100}_{\rm Pro}]}{C_{\rm Bin}[{\rm AUDS100}_{\rm Pro}]} = \frac{N_{\rm Bin}[{\rm AUDS100}]}{C_{\rm Bin}[{\rm AUDS100}]},
    	\end{equation}
    	which allows us to estimate the completeness in each bin in the AUDS100 sample. The number of galaxies in each bin of AUDS100$_{\rm Pro}$ and AUDS100 is shown in Figure \ref{Fig_09} (upper panel), with the error bars indicating the Poisson noise. The completeness for the AUDS100 sample is also shown in Figure \ref{Fig_09} (lower panel). In order to avoid small number statistics, we use a wide bin width of 2.0 dex, and remove the bins where either $N_{\rm Bin}[{\rm AUDS100}]$ or $N_{\rm Bin}[{\rm AUDS100}_{\rm Pro}]$ is smaller than 5. The completeness for both ${\rm AUDS100}_{\rm Pro}$ and AUDS100 is shown. We use Equation \ref{eq:sigmoid} to fit, and re-derive $C_2$ for the latter, as given in Table \ref{Tab_03}. We use the same method for the $Sub$ sample using galaxy numbers in $Sub_{\rm Pro}$ and $Sub$ instead. 
    	
    	As indicated above, the completeness should be zero for galaxies with low values of $(S_{\rm int}/\sigma)$ and high values of $W_{50}$ in Figure \ref{Fig_08}. To avoid the non-zero tail of the completeness function artificially affecting the effective volume, we truncate the completeness function (i.e.\ set to 0) at values of 0.03 and 0.35 for the AUDS100 and {\it Sub} samples, respectively. No galaxies are detected in this region.

	\subsubsection{Modified $V/V_{\rm max}$ test}\label{TcMethod}
        
        The modified $V/V_{\rm max}$ test was introduced by \citet{2001MNRAS.324...51R} to examine the completeness of optical catalogues. It can also be used to assess completeness of HI catalogues \citep{2004MNRAS.350.1210Z, 2016AJ....151...52S}. The test is based on the assumption of invariant shape of the HIMF, but is insensitive to large-scale structure (changes in density only). It is based on the data alone, and is therefore an ideal method to verify the completeness derived from our simulations.
        
        The noise level in our cubes varies at different spatial positions and frequencies. Hence, the noise level for each galaxy is different from each other. Following \citet{2015MNRAS.452.3726H}, and the simulations conducted above, we use the noise-scaled integrated flux $(S_{\rm int}/\sigma)$, and linewidth ($W_{50}$) to characterise the completeness. We construct $Z_0=\log(S_{\rm int}/\sigma)+\log(M_{\rm HI})$ in order to apply the $T_C$ method \citep{2001MNRAS.324...51R}.
        
    	\begin{figure}
	        \includegraphics[width=8cm,height=6cm]{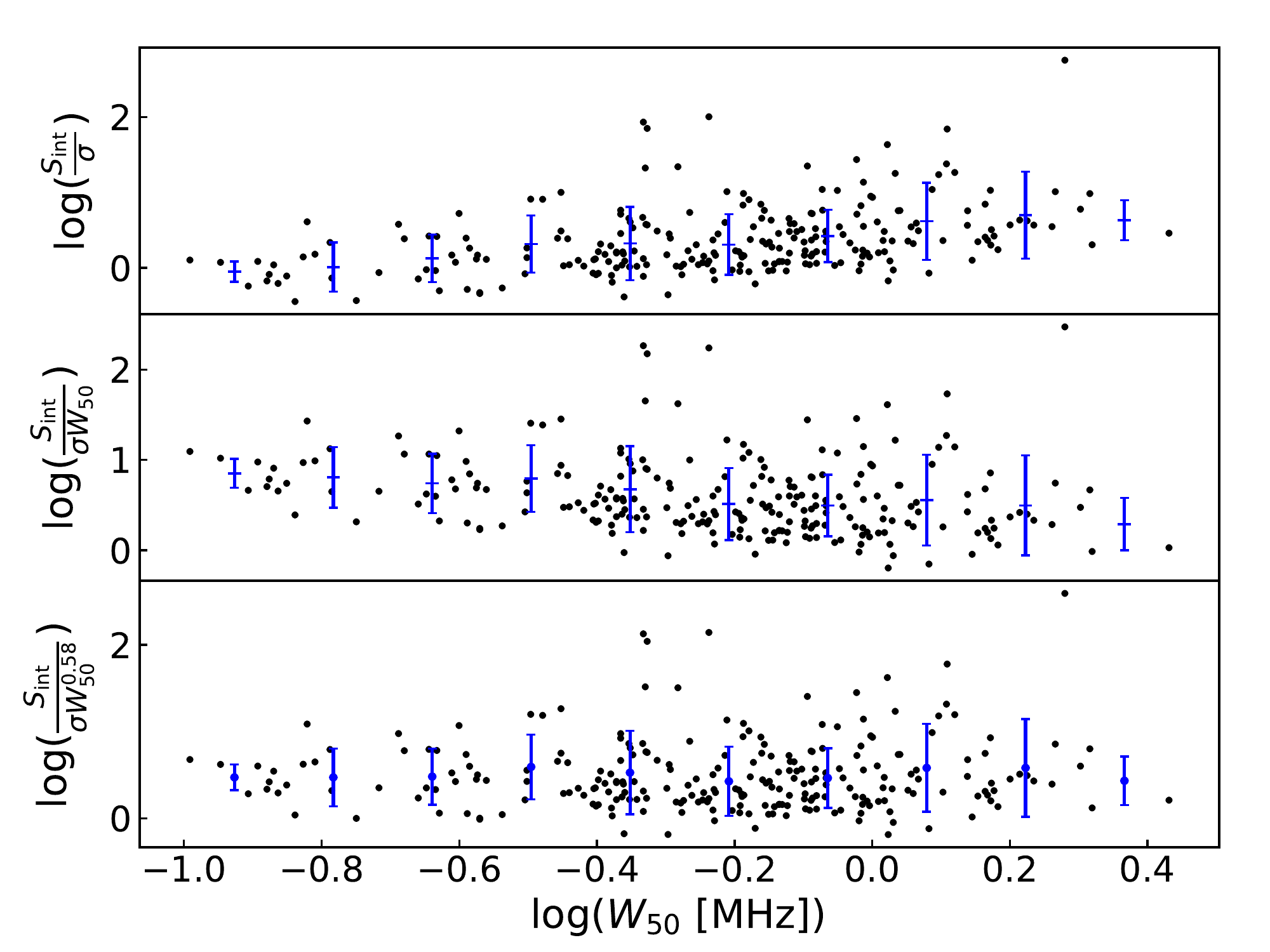}
	        \caption{The noise-scaled integrated flux, and scaled mean flux density for AUDS100 sample are shown in upper two panels. In the bottom panel, a velocity width power-law index of 0.58 was found to best fit the AUDS100 sample. For the {\it Sub} sample, a similar index 0.69 was found. $S_{\rm int}$, $\sigma$, $W_{50}$ are in the units of mJy~MHz, mJy, and MHz, respectively. }
	        \centering\label{Fig_10}
    	\end{figure}
    	
        In Figure \ref{Fig_10}, we show the distribution of noise-scaled integrated flux $(S_{\rm int}/\sigma)$, and noise-scaled mean flux density $(S_{\rm int}/\sigma W_{50})$ as a function of $W_{50}$. It clearly shows that the galaxies with narrow linewidths are easier to detect for a given integrated flux, and that galaxies with broad linewidth are easier to detect for a given mean flux density. The radiometer equation suggests a flat distribution should result by applying the scaling $W_{50}^\alpha$, with $\alpha=0.5$ for integrated flux and $W_{50}^{\alpha-1}$ for mean flux density. However, \citet{2016AJ....151...52S} suggest that baseline ripple makes galaxies with broad linewidth harder to detect, causing $\alpha$ to deviate from 0.5. The lower panel in Figure \ref{Fig_10} suggests $\alpha=0.58$ is appropriate for the AUDS100 sample. Similarly, $\alpha=0.69$ is derived for {\it Sub} sample. We therefore construct $Z_1=\log(S_{\rm int}/\sigma W_{50})+\log(M_{\rm HI})$, and $Z_2=\log(S_{\rm int}/\sigma W_{50}^\alpha)+\log(M_{\rm HI})$.
        
    	\begin{figure}
	        \includegraphics[width=8cm,height=6cm]{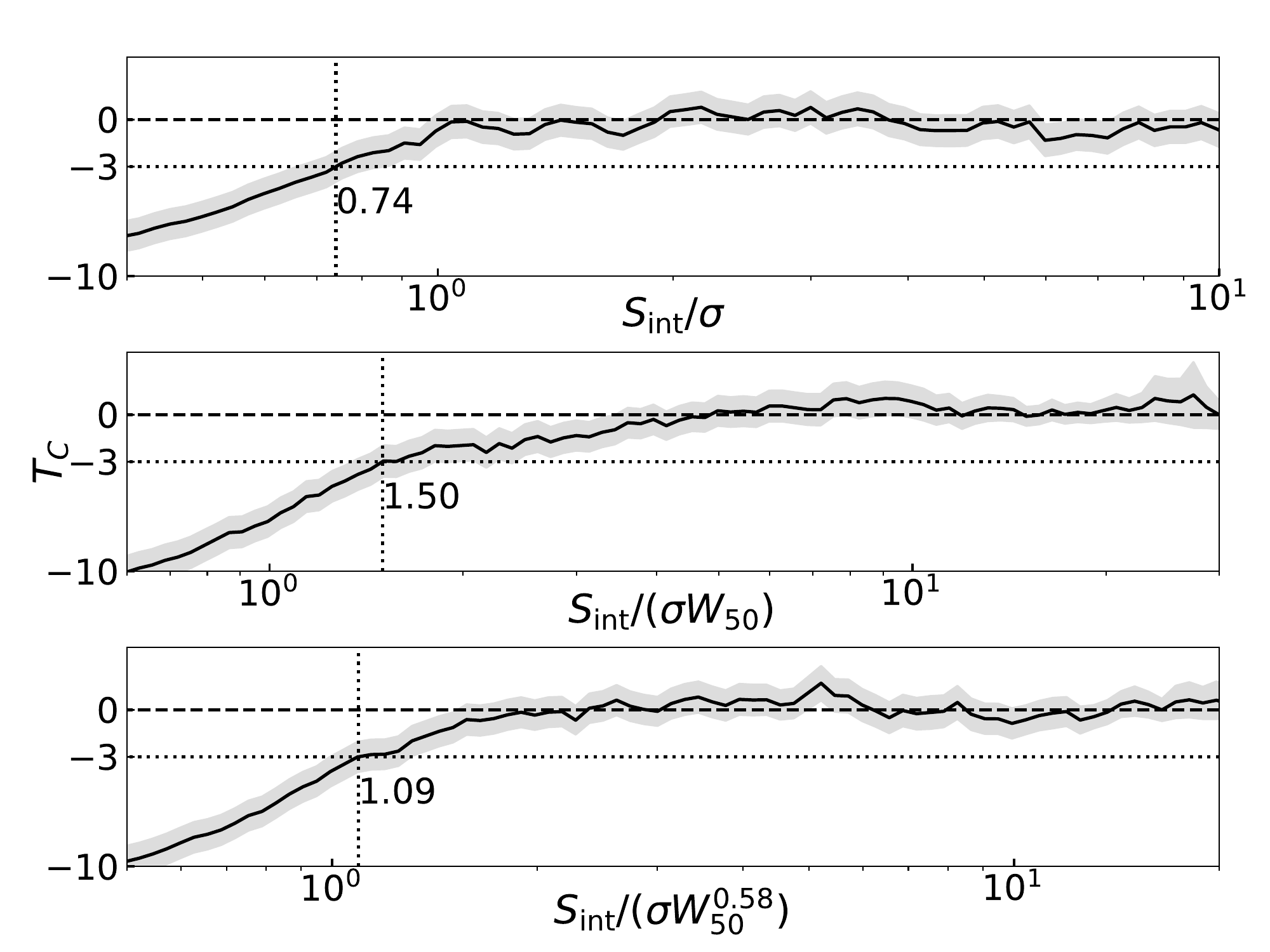}
	        \caption{The variation of the completeness statistic $T_C$, for $(S_{\rm int}/\sigma)$, $(S_{\rm int}/\sigma W_{50})$, and $(S_{\rm int}/\sigma W_{50}^\alpha)$. The units are mJy~MHz for $S_{\rm int}$, MHz for $W_{50}$, and mJy for $\sigma$. The grey area indicates errors due to counting statistics. The $3\sigma$ lower limit ($T_C=-3$) is shown by the dotted lines in each panel. The limit for $T_C=-3$ is also given in each panel by vertical dotted lines.}
	        \centering\label{Fig_11}
    	\end{figure}
    	
    	We show the results for the AUDS100 sample derived from the $T_C$ statistic in Figure \ref{Fig_11}. The sharpest drops to $T_C=-3$ occur at $(S_{\rm int}/\sigma)=0.74$ and $(S_{\rm int}/\sigma W_{50}^{0.58})=1.09$ in the first and third panels, respectively. This gives us 99.4 per cent confidence that the 100 per cent completeness limits are above these two values. $T_C$ drops more gradually for $(S_{\rm int}/\sigma W_{50})$, indicating that a single limit does not well characterise completeness for this parameter.

        We compare completeness from the $T_C$ test with our simulations in Figure \ref{Fig_05}, combining the two limits above in order to give the harshest constraints. The $T_C$ test gives a similar result to the simulation at a completeness level of 0.3 for galaxies with broad linewidths, and a similar dependence on linewidth. However, the $T_C$ test gives us a more strict constraint on galaxies with narrow linewidths, albeit this part of the plane is not as well populated. For the {\it Sub} sample, we derive $T_C=-3$ limits of $(S_{\rm int}/\sigma)=0.97$ and $(S_{\rm int}/\sigma W_{50}^{0.69})=1.61$, and a closer correspondence between the $T_C$ and simulation limits.

    \subsection{Reliability}\label{Reliability}
        
        AUDS100 sources were selected independently of SDSS, so cross-matching between the catalogues is a good measure of reliability. Combining SDSS and follow-up AAT spectroscopy, Section \ref{s:optical} shows that there are 97 (39 per cent) of AUDS100 sources which have optical counterparts. Unfortunately, the remaining AUDS100 sources do not have any spectroscopic observations. However, the photometric redshift analysis of Section \ref{s:optical} also shows that all of the remaining 150 (61 per cent) sources have plausible counterparts, though it is probable that $\sim$ three matches are unrelated. This suggests that reliability may be as high as 98 per cent, though this is likely an upper limit.
        
        \citet{2015MNRAS.452.3726H} measured reliability by placing synthetic galaxies in the AUDS60 data cube and compared the results of three by-eye source searchers. In this paper, we use negative detections in the final cubes to internally estimate reliability. The final cubes were inspected by eye for negative detections in the same manner as for positive detections (as described in Section \ref{SourceFinding}). 23 negative detections were found, mostly at the survey edge or in the strong RFI frequency range, and all had low SNR. The completeness parameters of these negative detections are small relative to AUDS100 sample. This indicates that AUDS100 galaxies which are: 1) located at the edge of survey area; 2) located in the strong RFI frequency range; or 3) have low SNR, are prone to being false detections. These 23 negative galaxies give an alternative estimate of AUDS100 reliability of $\sim$ 91 per cent.
        
        We have also formed a high-reliability, high-completeness sub-sample of 148 AUDS100  galaxies (`{\it Sub}') using the following criteria:
        \begin{itemize}
            \item Frequency greater than 1.25 GHz (to avoid RFI),
            \item Peak flux density greater than $2.5 \sigma$, where $\sigma$ is the local rms in mJy, and
            \item Integrated flux greater than $10.5 \sqrt{n} \sigma \Delta\nu$, where $n$ is number of channels over which the galaxy is detected and $\Delta\nu$ is channel width in MHz. 
        \end{itemize}
        None of the negative detections satisfy the above criteria, so the reliability for {\it Sub} should be very close to 100 per cent. In this high-reliability sub-sample, there are 83 (56 per cent) HI sources with spectroscopic redshifts from either SDSS or AUDSOC. This rate is close to that of HIPASS objects with optical spectroscopic redshifts (58 per cent) \citep{2005MNRAS.361...34D}, but higher than for the full AUDS sample (39 per cent).
        
\section{HI Mass Function}\label{Methods}
    
    The HI mass function (HIMF) is used to describe the average density of galaxies of different neutral hydrogen mass  in the Universe. Most studies \citep{2005MNRAS.359L..30Z, 2010ApJ...723.1359M, 2015MNRAS.452.3726H} parameterize the HIMF with a Schechter function \citep{1976ApJ...203..297S}:
    
    \begin{equation}
        \Phi(M_{\rm HI}) = \ln(10)\Phi_*(\frac{M_{\rm HI}}{M^*})^{\alpha+1}e^{-\frac{M_{\rm HI}}{M^*}},
    \end{equation}
    where $\alpha$ is the slope at low-mass end, $M^*$ is the characteristic mass, and $\Phi_*$ is the normalization constant. We adopt two methods to derive the HIMF, $\Sigma 1/V_{\rm max}$ and 2D Stepwise Maximum Likelihood (2DSWML). 
    
	\subsection{$\Sigma 1/V_{\rm max}$ method}\label{Method_1}
	    
	    In the $\Sigma 1/V_{\rm max}$ method \citep{1968ApJ...151..393S}, the maximum volume ($V_{\rm max}$) in which a galaxy can be detected is computed from its maximum detectable distance. The HIMF can then be derived, for example by summing $1/V_{\rm max}$ over different mass bins. It is a widely used method due to its speed and simplicity of implementation. However, it is sensitive to cosmic variance, which can bias shape and amplitude of the HIMF. For example, an incorrect slope can be caused by having a different local galaxy density (mainly low mass) as compared to with distant galaxies (mainly high mass) \citep{2010ApJ...723.1359M}. Here, we follow the procedure described in \citet{2015MNRAS.452.3726H}, in which effective volume ($V_{\rm eff}$) is used instead of $V_{\rm max}$.
	    
	    The method requires a noise and a volume cube for each field. For the noise cube, we use the method described in Section \ref{SourceFinding} to calculate the noise level at each pixel. The lower noise level is adopted in the overlap region between S0 and S1 bands. For the volume cube, the value represents the comoving volume for that pixel. The effective volume for each galaxy is computed from
	    
	    \begin{equation}
	    \label{Equ_07}
	        V_{\rm eff} = \Sigma V_{\rm co} \times f \times C_i ,
	    \end{equation}
	    where $V_{\rm co}$ is the comoving volume in each pixel, $f$ is the relative overdensity for correcting cosmic variance (see Section \ref{CosmicVariance}), and $C_i$ is the completeness for each galaxy (calculated from Equation \ref{eq:sigmoid}). The $V_{\rm eff}$ is the sum of all pixels in the noise and volume cubes.
	    
	    The HIMF is then derived by summing the inverse of the effective volume within each mass bin, with the error given by $\sqrt{\Sigma 1/V_{\rm eff}^2}$. A non-linear least squares (NLLS) method is then used to obtain the best-fit Schechter parameters. This method requires the binning of the data, which introduces sensitivity to the bin position and bin width. The effect is significant if the sample is small, such as for AUDS. The modified maximum likelihood (MML) method of \citet{2018MNRAS.474.5500O}, which can fit without binning, is therefore also adopted.
	    
	\subsection{2DSWML}
        
    	The 2DSWML method uses the likelihood of finding a galaxy with HI mass $M_{\rm HI, i}$ and velocity linewidth $W_{50,i}$ at distance $D_{i}$. As with $\Sigma 1/V_{\rm max}$, the method is model-independent, but is also robust against cosmic variance if the shape of the HIMF is invariant. It therefore provides an independent method for confirming the $\Sigma 1/V_{\rm max}$ results \citep{2005MNRAS.359L..30Z,2010ApJ...723.1359M}.
    	
    	The HI mass-linewidth plane is split into logarithmic mass and linewidth bins, and galaxies in each bin are counted. The local rms and distance of each galaxy is used to compute the completeness in the mass-linewidth plane for the galaxy, and the mean value of completeness in each bin is computed. The maximum likelihood solution was derived by iterating Equation B6 in \citet{2010ApJ...723.1359M} until the solution was stable. The errors are derived from inverse of its information matrix, as detailed in \citet{1988MNRAS.232..431E} and \citet{2000MNRAS.312..557L}. We apply the same constraint as used in \citet{2010ApJ...723.1359M},  
	
	    \begin{equation}
	        g= \Sigma_j \Sigma_k \left(\frac{M_{\rm HI,\it j}}{M_{\rm HI, ref}}\right)\left(\frac{W_{50, k}}{W_{50, \rm ref}}\right)\phi_{jk}\Delta m \Delta w - 1 = 0 ,
	    \end{equation}
	    where $m=\log(M_{\rm HI}~[h_{70}^{-2} \rm M_\odot])$, $w=\log(W_{50} [\rm MHz])$, $M_{\rm HI, \it j}$ and $W_{50, k}$ are the central HI mass and linewidth in bin {\it jk}, $\phi_{jk}$ is the stable solution in the $\log(M_{\rm HI})$-$\log(W_{50})$ plane, and $M_{\rm HI, ref}$ and $W_{50, \rm ref}$ are the mean mass and linewidth for the AUDS100 sample.
    	
    	To convert the probability density function into an HIMF, we need to evaluate the amplitude of the HIMF by matching the integral of the distribution to the inferred average density of galaxies in the survey volume. The selection function, $S(D)$, is used to normalize the survey volume and is given by following equation which gives the galaxy number weighted selection function:
    	
    	\begin{equation}
    	    S(D) = \frac{\int_w\int_m C_D(m, w, \sigma) \phi dm dw}{\int\int \phi dm dw}
    	\end{equation}
    	where $C_D(m, w, \sigma)$ is derived from Equ \ref{Equ_02} and \ref{eq:sigmoid}. The $\sigma$ is median value of noise at distance $D$. Considered the different noise at same distance in two survey field, we calculate selection function for each field. We adopt volume as weight to derive our final selection function. There are three estimators for the survey volume \citep{{1982ApJ...254..437D}}. We adopt the most stable estimator, $n_3$, to normalize $\phi_{jk}$. The HIMF is derived by integrating $\phi_{ij}$ over linewidth. The best fit (NLLS) parameters were then derived for the Schechter function.
    	
    \subsection{Cosmic variance} \label{CosmicVariance}
    
        Cosmic variance is a statistical uncertainty in estimating the number density of galaxies caused by the underlying large-scale density fluctuations. The uncertainty is in particularly significant for high-redshift, deep pencil-beam surveys. Methods for correcting for the influence of cosmic variance on the HIMF include: (1) using an optical catalogue to measure the relative density distribution \citep{2015MNRAS.452.3726H,2016MNRAS.457.4393J}; and (2) using a density distribution model \citep{2010ApJ...723.1359M}. We adopt the first method to correct for the cosmic variance, calculating the overdensity for a given redshift from: 
        \begin{equation}
            f(z)=\frac{\rho(z)}{\rho_{\rm ref}(z)}, 
        \end{equation}
        where $\rho(z)$ and $\rho_{\rm ref}(z)$ are number density of galaxies in each of the AUDS field and a reference field, respectively. 

\begin{figure}
            \includegraphics[width=8cm,height=5cm]{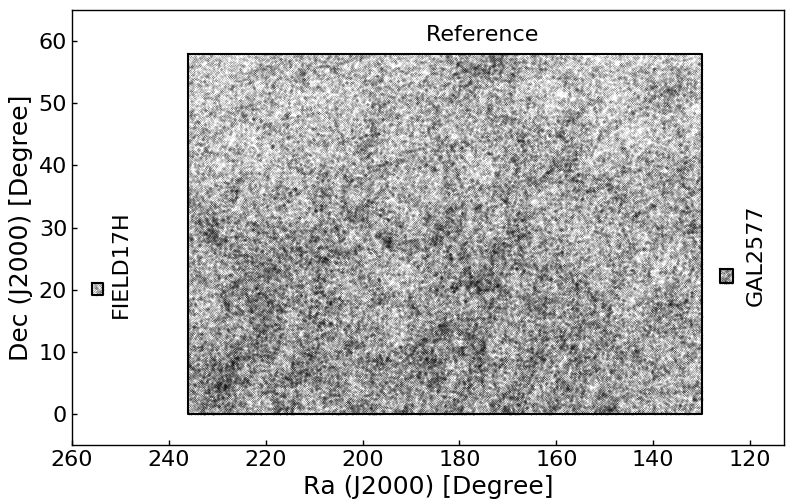}
            \caption{Projected number distribution of late-type  SDSS galaxies (as described in Section \ref{CosmicVariance}) in the two AUDS fields (FIELD17H and GAL2577), and the reference volume.}
            \centering\label{Fig_12}
        \end{figure}

        To calculate the number density of galaxies in these fields, we select late-type galaxies from the SDSS DR14 catalogue \citep{2018ApJS..235...42A}. We preferentially select late-type galaxies using the criteria: $u^*-r^*< 2.22$, where $u^*$ and $r^*$ are model magnitudes as described in \citet{2001AJ....122.1861S}, and also select galaxies with redshift uncertainty $\sigma_z$< 0.01. Late-type galaxies better represent the galaxy population in an HI survey \citep{2018MNRAS.477....2J}. Figure~\ref{Fig_12} shows the selected galaxies for FIELD17H (253.5$^\circ$ < $\alpha$ < 255.8$^\circ$, +19.2$^\circ$< $\delta$ < +21.1$^\circ$), GAL2577 (123.6$^\circ$< $\alpha$ < 126.3$^\circ$, +21.0$^\circ$< $\delta$ < +23.3$^\circ$) and the much larger SDSS reference field (130.0$^\circ$ < $\alpha$ < 236.0$^\circ$, 0.0$^\circ$ < $\delta$ < +58.0$^\circ$). The selected reference field is the same as in the cosmic variance study of \citet{2010MNRAS.407.2131D}, for which  the volume is $6.14 \times 10^7$ Mpc$^3$ out to $z=0.17$. 
    
\begin{figure}
            \includegraphics[width=8cm,height=5cm]{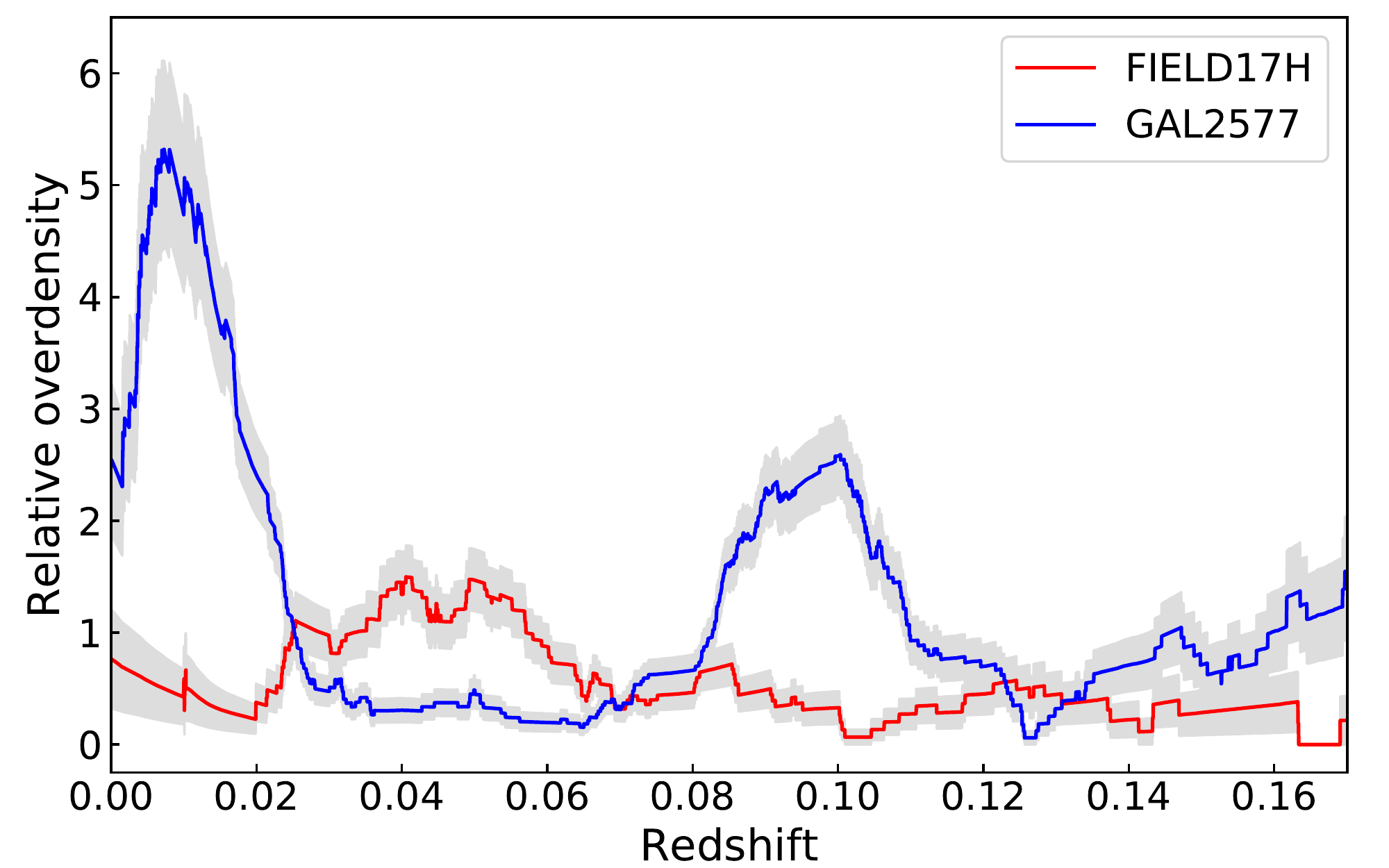}
            \caption{Relative overdensities are derived from blue SDSS galaxies. The figure shows it as a function of redshift in the two AUDS fields, FIELD17H and GAL2577, in redshift bins of width 0.02. The shaded grey represents the Poisson error.}
            \centering\label{Fig_13}
        \end{figure}

        For redshift bin of width 0.02, we calculate the ratio of the density of galaxies in AUDS and the reference field up to a redshift of 0.17. (see Figure~\ref{Fig_13}). The error in the measured overdensity is calculated from  Poisson statistics. Two peaks with overdensities over 2 are seen around redshift 0.01 and 0.09 in the GAL2577 field. These peaks are also consistent with HI overdensities.

\section{Results}\label{HIMF}
    
    \begin{figure*}
        \includegraphics[width=12cm,height=9cm]{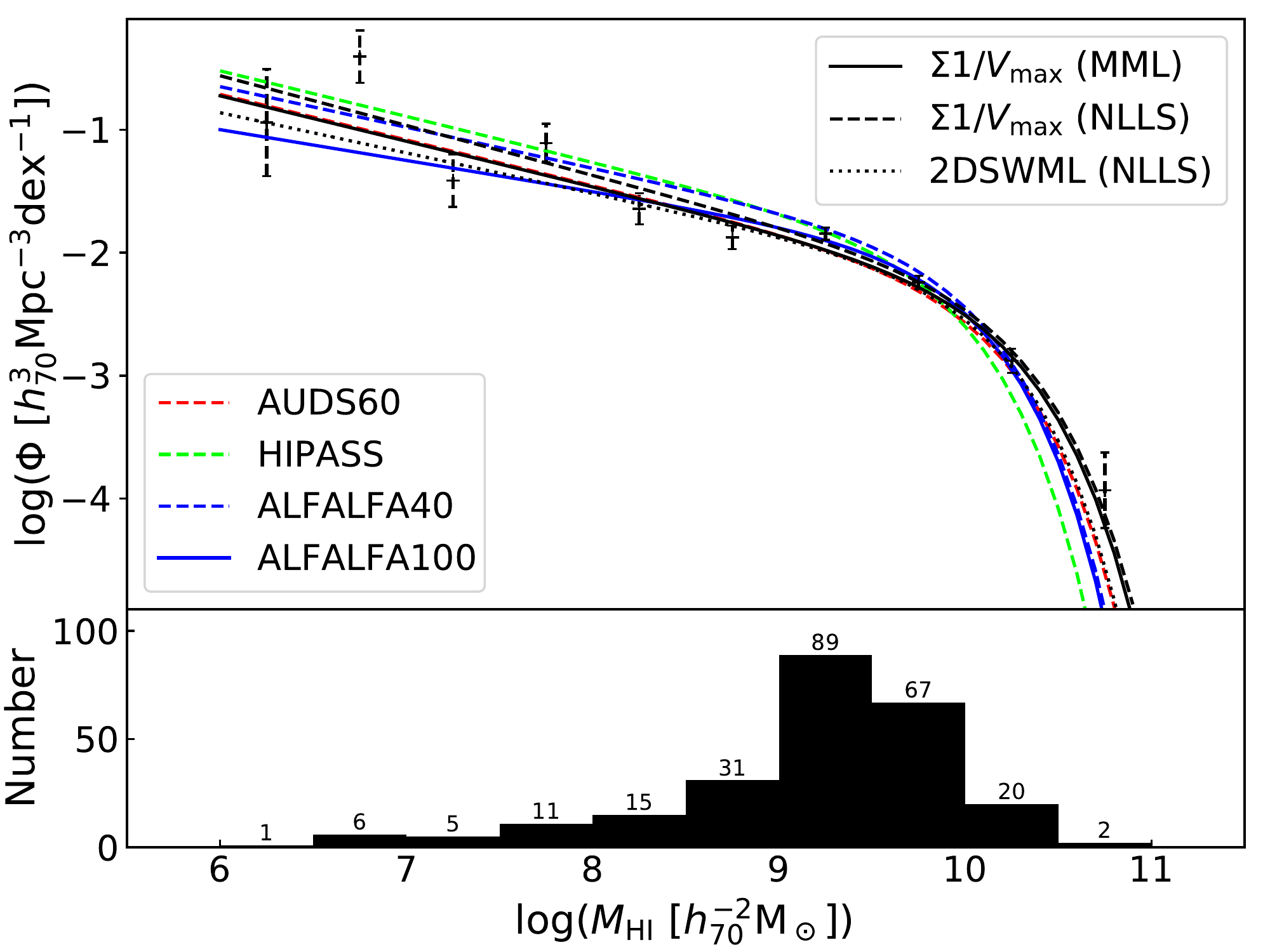}
        \caption{The HIMF derived using the $\Sigma 1/V_{\rm max}$ and 2DSWML methods is shown in the upper panel. Two different methods were used to fit smooth Schechter functions to the $\Sigma 1/V_{\rm max}$ data: modified maximum likelihood (MML) (solid black line); and non-linear least square (NLLS) (dashed black line). The results from previous work is overlaid. The distribution of HI masses in AUDS100 is shown in the lower panel.}
        \centering\label{Fig_14}
    \end{figure*}
    
    \begin{table*}
        \centering
        \caption{The best-fit parameters for a Schechter function parameterization of the HIMF for AUDS100 and sub-samples of AUDS100. Two different methods are used for computing the HIMF, and two different methods are employed for fitting the Schechter function. The {\it Ref} sample removes 10 low-mass and high-mass galaxies from AUDS100 (see Section~\ref{HIMF}), and the {\it Sub} sample consists of the 148 galaxies with high reliability and completeness (see Section~\ref{Reliability}).}
        \begin{tabular}{rrrrrrrr}
            \hline
            Sample  &  Method      & Fitting &  $\Phi_* [10^{-3} h_{70}^3 {\rm Mpc}^{-3}]$ & $\log(M^*~[h_{70}^{-2}{\rm M}_\odot])$ & $\alpha$    & $\Omega_{\rm HI}({\rm int})~[10^{-4} h_{70}^{-1}]$  & $\Omega_{\rm HI}({\rm sum})~[10^{-4} h_{70}^{-1}]$\\
                \hline
            AUDS100    & $1/V_{\rm max}$ & MML  & 2.41(0.57)  & 10.15(0.09) & $-1.37$(0.05) & 3.55(0.30) & ...\\
            AUDS100    & $1/V_{\rm max}$ & NLLS & 2.46(0.55)  & 10.17(0.08) & $-1.40$(0.05) & 4.04(0.34) & 3.93(0.68)\\
            AUDS100    & 2DSWML          & NLLS & 2.81(0.55)  & 10.06(0.06) & $-1.33$(0.05) & 3.18(0.25) & 3.09(0.55)\\
            {\it Ref}  & $1/V_{\rm max}$ & MML  & 3.82(0.87)  & 9.98(0.08)  & $-1.25$(0.07) & 3.27(0.26) & 3.60(0.65)\\
            {\it Sub}  & $1/V_{\rm max}$ & MML  & 4.56(1.23)  & 9.91(0.09)  & $-1.25(0.08)$ & 3.33(0.32) & 3.72(0.71)\\
            FIELD17H   & $1/V_{\rm max}$ & MML  & 2.22(0.98)  & 10.33(0.17) & $-1.38$(0.09) & 5.00(0.77) & 5.27(1.15)\\
            GAL2577    & $1/V_{\rm max}$ & MML  & 2.76(0.78)  & 10.03(0.10) & $-1.37$(0.06) & 3.08(0.30) & 3.36(0.64)\\
            Low-$z$    & $1/V_{\rm max}$ & MML  & 3.32(1.19)  & 9.92(0.15)  & $-1.37$(0.07) & 2.94(0.43) & 3.07(0.64)\\
            High-$z$   & $1/V_{\rm max}$ & MML  & 2.81(1.46)  & 10.13(0.15) & $-1.36$(0.21) & 3.91(0.38) & 3.51(0.75)\\
            \hline
        \end{tabular}
        \label{Tab_04}
    \end{table*}
        
    \begin{table*}
        \centering
        \caption{The best-fit parameters for a Schechter function parameterization of the HIMF for different surveys, including HIPASS, ALFALFA, and AUDS.}
        \begin{tabular}{rrrrrrrrr}
            \hline
            Reference   &  Sample  &  Method      & $\Phi_* [10^{-3} h_{70}^3 {\rm Mpc}^{-3}]$ & $\log(M^*~[h_{70}^{-2}{\rm M}_\odot])$ & $\alpha$    & $\Omega_{\rm HI}~[10^{-4} h_{70}^{-1}]$\\
            \hline
            This work                    & AUDS100    & $1/V_{\rm max}$ & 2.41(0.57)  & 10.15(0.09) & $-1.37$(0.05) & 3.55(0.30) \\
            \citet{2015MNRAS.452.3726H}  & AUDS60     & $1/V_{\rm max}$ & 2.65(0.48)  & 10.10(0.04) & $-1.37$(0.03) & 3.33(0.10) \\
            \citet{2018MNRAS.477....2J}  & ALFALFA100 & $1/V_{\rm max}$ & 4.5(0.2) & 9.94(0.01) & $-1.25$(0.02) & 3.9(0.1) \\
            \citet{2010ApJ...723.1359M}  & ALFALFA40  & 2DSWML          & 4.8(0.3) & 9.96(0.02) & $-1.33$(0.02) & 4.3(0.3) \\
            \citet{2005MNRAS.359L..30Z}  & HIPASS     & 2DSWML          & 4.88(0.65) & 9.86(0.03) & $-1.37$(0.03) & 3.75(0.43) \\
            \hline
        \end{tabular}
        \label{Tab_05}
    \end{table*}
        
    The HIMFs derived from the $\Sigma 1/V_{\rm max}$ and 2DSWML methods are presented in Figure \ref{Fig_14}, together with the HI mass distribution for the AUDS100 sample. The binned and unbinned data from the $\Sigma 1/V_{\rm max}$ method are fitted with the NLLS and MML algorithms, respectively. Since MML cannot be applied to the binned output of 2DSWML, only the NLLS algorithm is used. Mass bins below $10^{7.5}~h_{70}^{-2} {\rm M}_\odot$ and above $10^{10.5}~h_{70}^{-2} {\rm M}_\odot$ generally have larger errors due to small number statistics. We present the best fit parameters for each fitted Schechter function in Table \ref{Tab_04}. The derived parameters from both fitting methods are consistent within the error. For the $\Sigma 1/V_{\rm max}$ method, we find that the bin width affects the parameters derived using the NLLS fitting. The effect is prominent when the sample size is small in some bins. On the other hand, MML does not use bins, and takes into account the error in the masses and is generally a more robust algorithm. Comparing the HIMFs derived from the MML fitted $\Sigma 1/V_{\rm max}$ and the NLLS fitted 2DSWML, we find that the derived parameters are consistent within errors. We adopt the result from the MML fitted $\Sigma 1/V_{\rm max}$ method for the rest of the paper. 
        
    We also compare with the result from AUDS60 \citep{2015MNRAS.452.3726H} and find a similar value for $\alpha$, but a smaller $M^*$ and a larger $\Phi_*$, though all within the combined errors. The difference is that we have included galaxies with $z>0.13$, which were discarded in AUDS60 due to strong RFI. This gives us a more distant volume, which in turn influences the calculation of effective volume for high mass galaxies.
        
    Figure \ref{Fig_14} also compares with the results from HIPASS \citep{2005MNRAS.359L..30Z}, ALFALFA40 \citep{2010ApJ...723.1359M}, and ALFALFA100 \citep{2018MNRAS.477....2J}. Our result shows a good consistency with HIPASS and ALFALFA40 at the low-mass end. However, ALFALFA100 gives a flatter slope, $\alpha = -1.25$. The knee mass $M^*$ in AUDS100 is larger than in HIPASS and ALFALFA, but still consistent at the 2$\sigma$ level with ALFALFA40 and ALFALFA100. For the normalization parameters $\Phi_*$, \citet{2005MNRAS.359L..30Z} measured $\Phi_* = 4.88 \times 10^{-3} h_{70}^3 \rm Mpc^{-3}$ for HIPASS, and \citet{2010ApJ...723.1359M} and \citet{2018MNRAS.477....2J} measured $\Phi_* = 4.8 \times 10^{-3} h_{70}^3 \rm Mpc^{-3}$ and $4.5 \times 10^{-3} h_{70}^3 \rm Mpc^{-3}$ for ALFALFA40 and ALFALFA100, respectively. The AUDS100 value of $\Phi_* = 2.41 \times 10^{-3} h_{70}^3 \rm Mpc^{-3}$ is significantly smaller than the results from both HIPASS and ALFALFA.
        
    \begin{figure}
        \includegraphics[width=8cm,height=6cm]{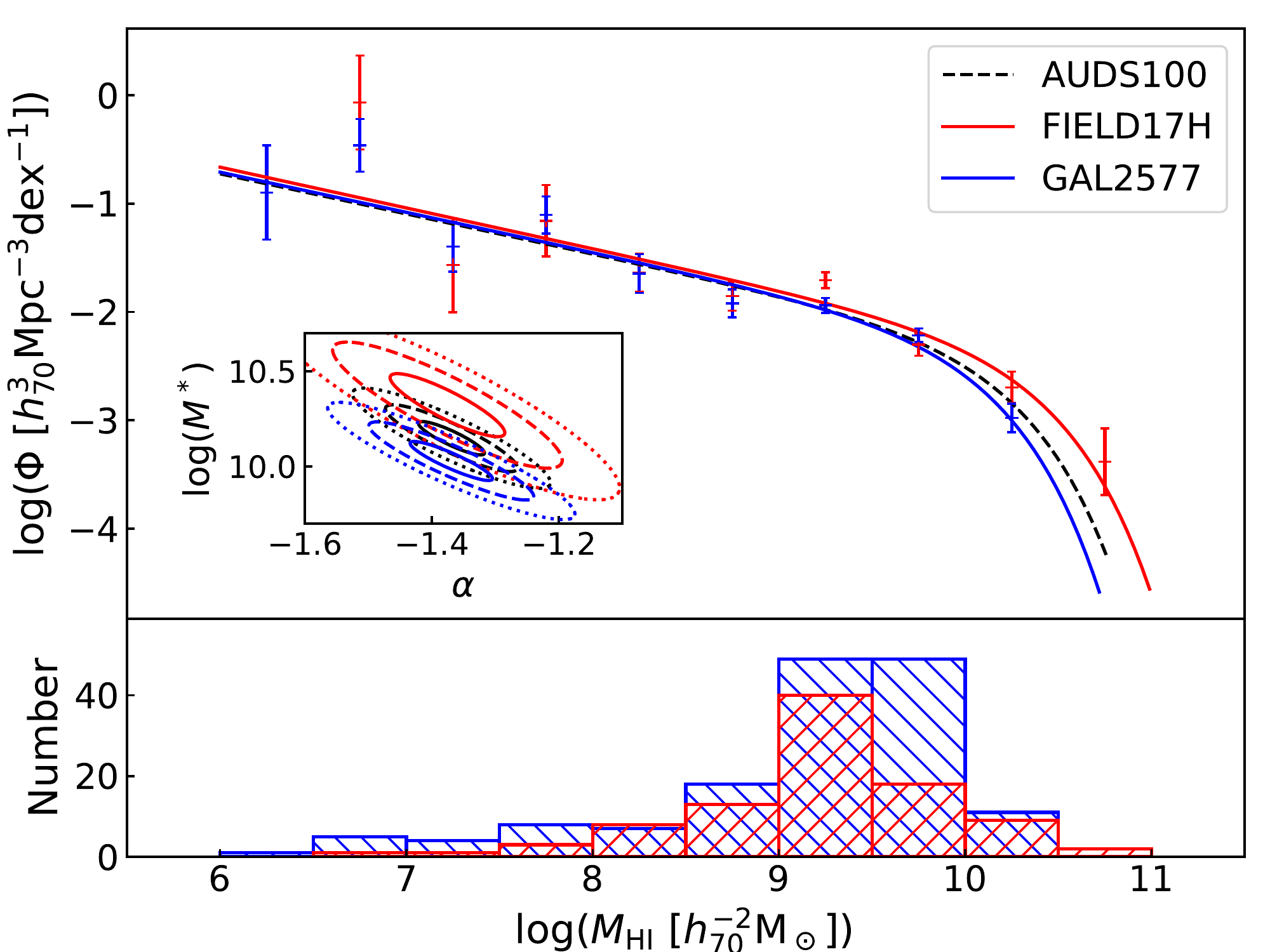}
        \caption{We show the distribution of galaxy mass in two fields in the lower panel. The red and blue hatched bars represent FIELD17H and GAL2577, respectively. HIMF for both field is shown in upper panel. We also show the error ellipse in $\alpha$-$\log(M^*)$ plane, where $M^*$ is in units of $h_{70}^{-2}{\rm M_\odot}$.}
        \centering\label{Fig_15}
    \end{figure}
    
    To investigate if the lower normalization reflects residual cosmic variance, even after the corrections discussed in Section~\ref{CosmicVariance}, we separately fit the two fields, FIELD17H and GAL2577. These fields are located in opposite parts of the sky (Figure \ref{Fig_12}), and differ in their SDSS spectroscopic redshift density (80 and 185 late-type galaxies, respectively) and HI source density (95 and 152, respectively). Table \ref{Tab_04} and Figure \ref{Fig_15} indicate that the two fields in fact have similar values for $\alpha$ and $\Phi_*$, but that the lower density FIELD17H has a slightly larger $M^*$ with 1$\sigma$ significance. Considering the possible influence from the slightly different mass ranges in the two sub-samples, we narrow the mass range to 10$^{6.6}$ - 10$^{10.3}$ $h_{70}^{-2} {\rm M}_\odot$, giving 89 and 149 galaxies in FIELD17H and GAL2577, respectively. The MML method gives Schechter parameters $\log(M^* [h_{70}^{-2} {\rm M}_\odot]) = 9.85 \pm 0.13$ and $\alpha = -1.18 \pm 0.13$ in FIELD17H, and $\log(M^* [h_{70}^{-2} {\rm M}_\odot]) = 9.97 \pm 0.10$ and $\alpha = -1.33 \pm 0.07$ in GAL2577. Both $M^*$ and $\alpha$ are consistent with each other within 1$\sigma$. There are numerous studies of the variation of the HIMF with environment (see Section 1). However, the overall effects of environment on the HIMF are currently not well understood and the results are not consistent between studies. Wide field surveys such as WALLABY \citep{2020Ap&SS.365..118K} will be required to further examine the role of environment on the HIMF.
    
    
    
    \begin{figure}
        \includegraphics[width=8cm,height=6cm]{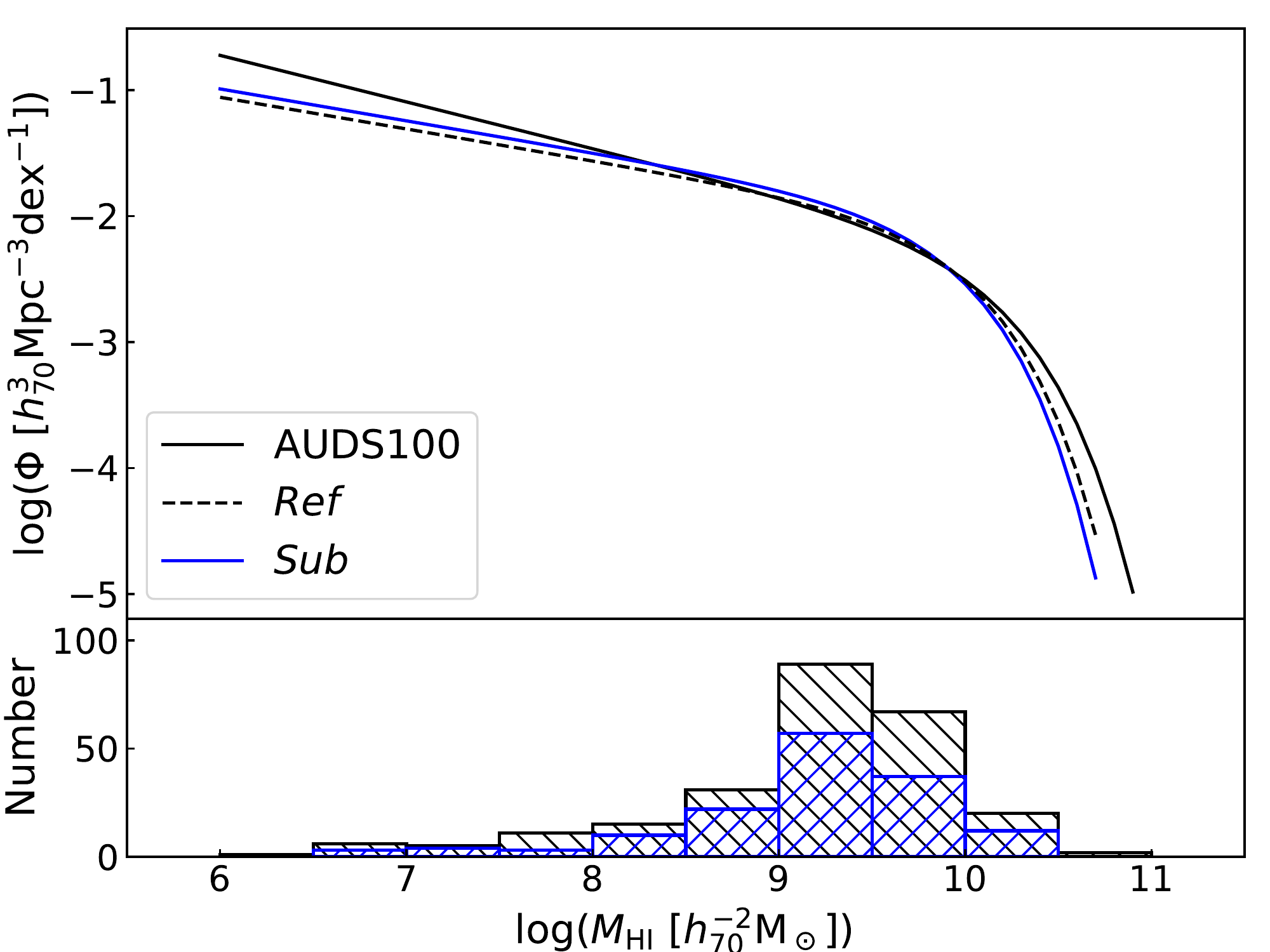}
        \caption{The HI mass distribution of galaxies in both the AUDS100 and {\it Sub} samples is shown in the lower panel by the black and blue histograms, respectively. The HIMF from AUDS100, {\it Sub}, and an AUDS100 sub-sample  (indicated by {\it Ref}) which excludes low and high-mass galaxies ($7.0 < \log({M_{\rm HI}}~[h_{70}^{-2}M_\odot]) < 10.5$) is shown in the upper panel.}
        \centering\label{Fig_16}
    \end{figure}
    
    To assess the effects of completeness and reliability, we also calculate the HIMF using the high reliability and high completeness {\it Sub} sample discussed in Section \ref{Reliability}. This gives 148 galaxies as compared to 247 galaxies in AUDS100. Figure~\ref{Fig_16} compares the two HIMFs as as fitted with the MML algorithm ($\Sigma 1/V_{\rm max}$ method) to derive the Schechter function parameters (see Table \ref{Tab_04}). There appears to be a deficit of low and high-mass galaxies, reflected in a less steep faint-end slope and smaller value for $M^*$, respectively, but a larger value of $\Phi_{\rm *}$. The difference is probably due to the lack of galaxies at both mass ends of the mass distribution in the {\it Sub} sample (see bottom panel of Figure~\ref{Fig_16}). If we exclude the 10 AUDS100 galaxies  outside of the range $7.0 < \log({\rm M_{HI}}~[h_{70}^{-2}{\rm M}_\odot]) <10.5$ (indicated by {\it Ref}) and re-derive the HIMF, the results become consistent (see Figure~\ref{Fig_16}).
    
    \begin{figure}
        \includegraphics[width=8cm,height=6cm]{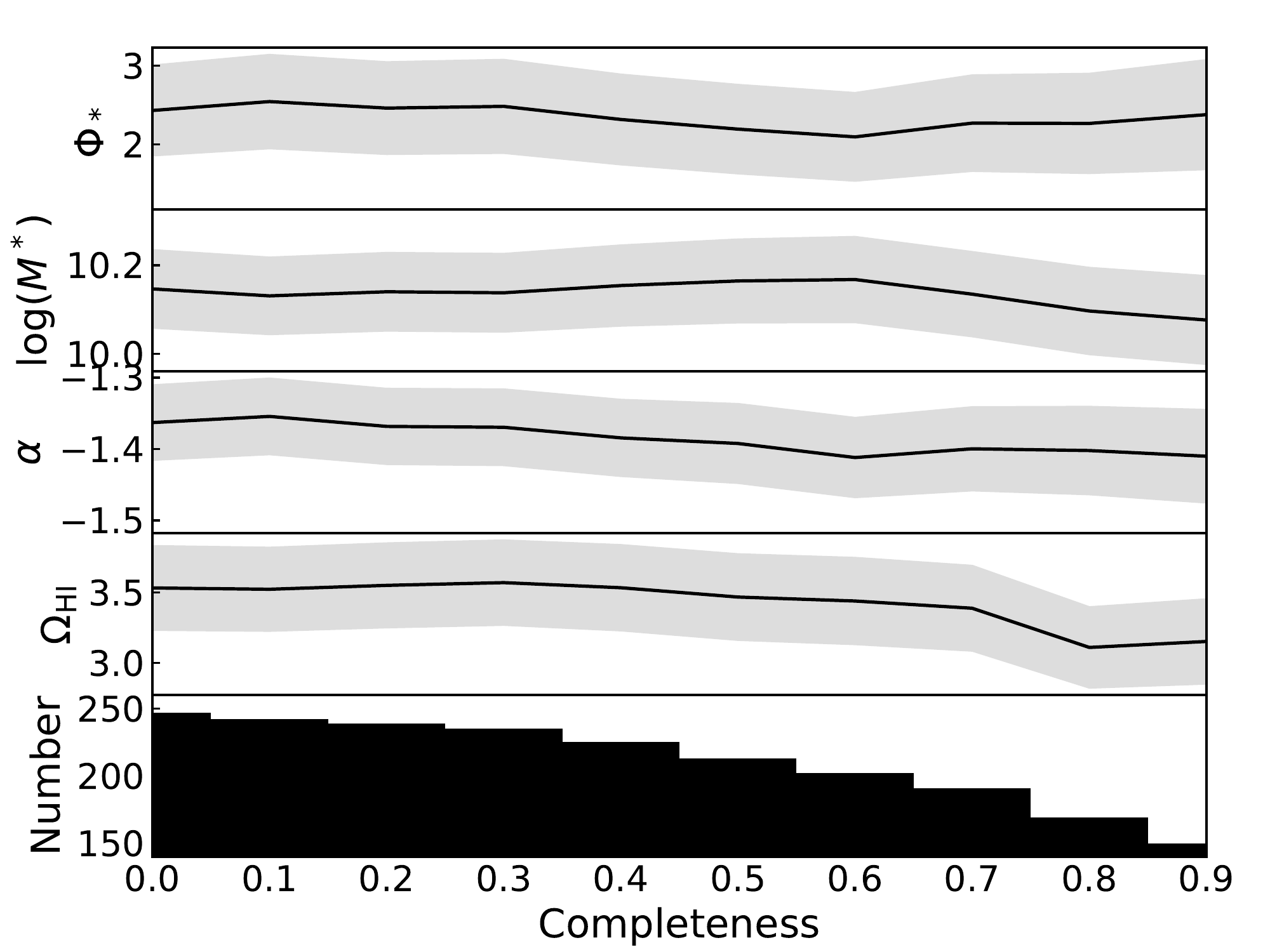}
        \caption{The variation of $\Phi_*~[10^{-3} h_{70}^3 {\rm Mpc}^{-3}]$, $\log(M^*~[h_{70}^{-2}{\rm M}_\odot])$, $\alpha$, and $\Omega_{\rm HI}~[10^{-4} h^{-1}]$ with different completeness cutoffs. The bottom panel shows the number of galaxies above a given completeness level. The grey regions represent the 68 per cent confidence intervals.}
        \centering\label{Fig_17}
    \end{figure}
    
    To further assess the effect of completeness on the HIMF, we use sub-samples with different completeness cutoffs, then re-compute and re-fit the HIMF. The results are shown in Figure \ref{Fig_17}. No substantial variation of HIMF parameters is seen, although the slope $\alpha$ appears to gradually steepen with increasingly complete sub-samples. At all completeness levels, the effective volume is corrected for completeness using Equation~\ref{Equ_07}, so the lack of substantial change in the HIMF parameters is an indication that our completeness corrections are reasonably accurate.
    
\section{Discussion}\label{Discussion}

    \subsection{Cosmic HI Density $\Omega_{\rm HI}$}\label{OmegaHI}
        
        HI mass density ($\rho_{\rm HI}$) is derived by integrating the mass-weighted HIMF. It can be approximated with the analytic expression:
        
        \begin{equation}
            \rho_{\rm HI} = \Gamma(\alpha + 2) \Phi_* M^*
        \end{equation}
        in which $\Gamma$ is Euler Gamma function, and $\Phi_*$, $M^*$ and $\alpha$ are the best fitted parameters of the Schechter function. The cosmic HI density is then given by:
        
        \begin{equation}
            \Omega_{\rm HI} = \frac{\rho_{\rm HI}}{\rho_{{\rm crit}(z=0)}} = \frac{8 \pi G}{3 H_0^2} \rho_{\rm HI}
        \end{equation}
        where $G$ is gravitational constant, $\rm 4.30 \times 10^{-9}~Mpc~{\rm M}_\odot^{-1}~(km~s^{-1})^2$, and $H_0$ is Hubble constant, $\rm 70~{\it h_{70}}~km~s^{-1}~Mpc^{-1}$ at $z=0$. We also compute $\Omega_{\rm HI}$ in a direct method by summing up the mass weighted HIMF data points. Taking into account the error of $\log(M_{\rm}~[h_{70}^{-2} \rm M_\odot])$, $\sqrt{\Delta\log(M_{\rm HI}~[h_{70}^{-2} \rm M_\odot])^2/12}$, $\Omega_{\rm HI}(\rm sum)$ has larger uncertainty than $\Omega_{\rm HI}(\rm int)$. We list the computed $\Omega_{\rm HI}$ for different samples in Table \ref{Tab_04}. We find $\Omega_{\rm HI}(\rm int) = 3.55(0.30) \times 10^{-4} h_{70}^{-1}$ and $\Omega_{\rm HI}(\rm sum) = 3.93(0.68) \times 10^{-4} h_{70}^{-1}$ for AUDS100, which is reasonably consistent with results from HIPASS ($3.75(0.43) \times 10^{-4} h_{70}^{-1}$) and ALFALFA ($3.9(0.1) \times 10^{-4} h_{70}^{-1}$) within a 1-$\sigma$ uncertainty. 
        
        \begin{figure}
            \includegraphics[width=8cm,height=5cm]{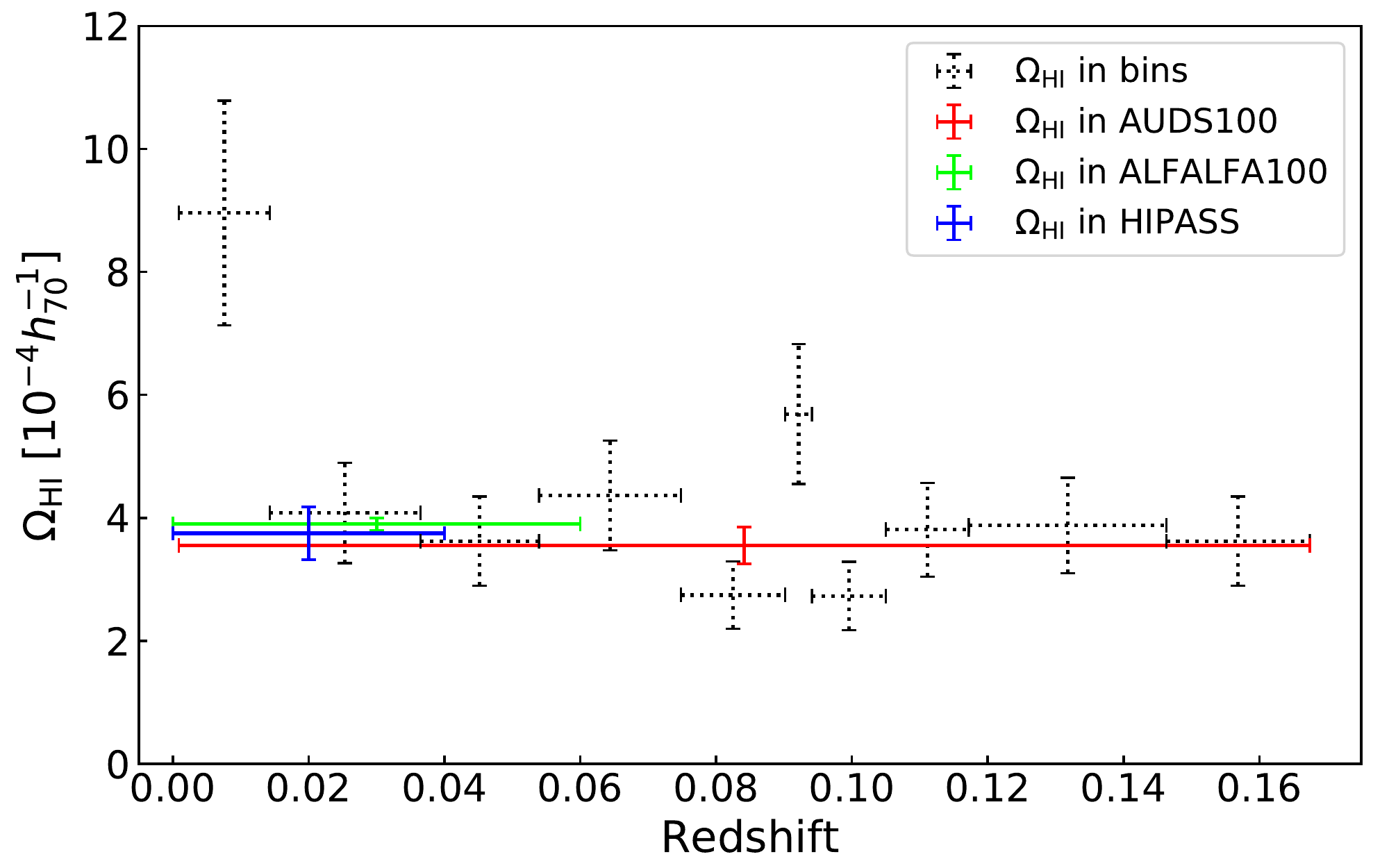}
            \caption{The cosmic HI density $\Omega_{\rm HI}$ in different redshift bins. The horizontal error bars indicate the redshift span, while the vertical error bars show $1\sigma$ uncertainties for each redshift bin. The black dotted symbols are the results for the individual redshift bins, for which the galaxy numbers are equal. The result for the whole AUDS100 sample is shown by the red symbol and error bar. We also show the result from HIPASS and ALFALFA by blue and green symbols, respectively.}
            \centering\label{Fig_18}
        \end{figure}

	\subsection{Evolution}\label{HIMFEvolution}
	    
        The sensitivity of AUDS makes it suitable for beginning to explore the evolution of $\Omega_{\rm HI}$ within the range covered by the ALFA receiver. We divide AUDS into ten redshift bins containing equal numbers of galaxies. Due to sensitivity and cosmic variance, the span for each redshift bin is not the same. We also adopt the characteristic mass and low-mass slope of the full AUDS100 sample, and only fit for the normalization $\Phi_*$ in the different redshift bins. The results are shown in Figure \ref{Fig_18}. There is no clear trend with redshift. This is consistent with the the prediction from cosmological simulations in \citet{2014MNRAS.440..920L, 2015MNRAS.453.2315K, 2017MNRAS.467..115D}, in which the $\Omega_{\rm HI}$ shows little evolution for $z < 2$. The upturn in the lowest-redshift bin may be an indication of existence of an `ankle' at the low-mass end of HIMF, rather than a change in overall density. Most of the galaxies in this bin have $\log(M_{\rm HI}~[h_{70}^{-2}{\rm M}_\odot]) < 8$.
        
        \begin{figure*}
            \includegraphics[width=16cm,height=7.27cm]{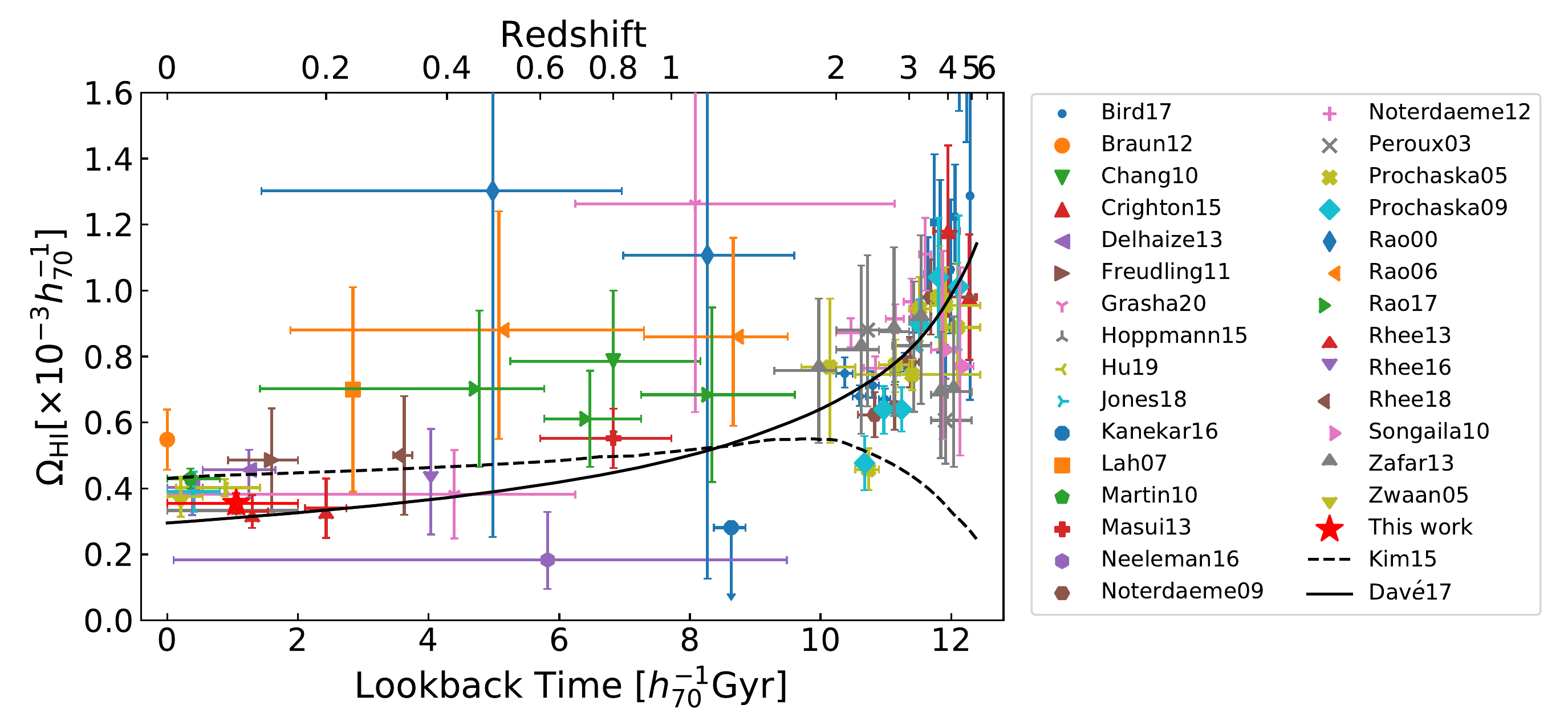}
            \caption{Measurements of $\Omega_{\rm HI}$ using different techniques: (1) HI sources detected using the 21-cm emission line: \citet{2005MNRAS.359L..30Z}, \citet{2010ApJ...723.1359M}, \citet{2011ApJ...727...40F}, \citet{2015MNRAS.452.3726H}, \citet{2018MNRAS.477....2J}; (2) HI spectral stacking: \citet{2007MNRAS.376.1357L}, \citet{2013MNRAS.433.1398D}, \citet{2013MNRAS.435.2693R}, \citet{2016ApJ...818L..28K}, \citet{2016MNRAS.460.2675R}, \citet{2018MNRAS.473.1879R}, \citet{2019MNRAS.489.1619H}; (3) HI intensity mapping: \citet{2010Natur.466..463C}, \citet{2013ApJ...763L..20M}; and (4) damped Lyman-$\alpha$ (DLA): \citet{2000ApJS..130....1R}, \citet{2003MNRAS.346.1103P}, \citet{2005ApJ...635..123P}, \citet{2006ApJ...636..610R}, \citet{2009A&A...505.1087N}, \citet{2009ApJ...696.1543P}, \citet{2010ApJ...721.1448S}, \citet{2012ApJ...749...87B}, \citet{2012A&A...547L...1N}, \citet{2013A&A...556A.141Z}, \citet{2015MNRAS.452..217C}, \citet{2016ApJ...818..113N}, \citet{2017MNRAS.466.2111B}, \citet{2017MNRAS.471.3428R}, \citet{2020MNRAS.498..883G}. The predictions from semi-analytic model \citep{2015MNRAS.453.2315K} and hydrodynamic simulation {\sc MUFASA} \citep{2017MNRAS.467..115D} are shown by the black dashed line and the black solid line, respectively.}
            \centering\label{Fig_19}
        \end{figure*}
        
        Several techniques are available for measuring $\Omega_{\rm HI}$ at different redshifts, which gives an insight into galaxy evolution. Detecting HI in individual galaxies in the 21-cm emission line is the main technique at low redshift (e.g. \citealp{2015MNRAS.452.3726H, 2018MNRAS.477....2J}). Due to tremendous observing time required beyond the local Universe, HI spectral stacking \citep{2018MNRAS.473.1879R, 2019MNRAS.489.1619H} and HI intensity mapping \citep{2010Natur.466..463C,2013ApJ...763L..20M} techniques are preferred at intermediate redshift. For high-redshift galaxies, damped Lyman $\alpha$ (DLA) absorption lines has been the main method to derive $\Omega_{\rm HI}$ \citep{2017MNRAS.471.3428R, 2020MNRAS.498..883G} (21-cm absorption line measurements are also possible, but have a temperature dependence).
        
        We show selected measurements of $\Omega_{\rm HI}$ using the above techniques in Figure \ref{Fig_19}, together with predictions from a semi-analytic model \citep{2015MNRAS.453.2315K} and a hydrodynamic simulation ({\sc MUFASA}) \citep{2017MNRAS.467..115D}. All the measurements of $\Omega_{\rm HI}$ from previous studies are converted to flat cosmology with $H_0 = 70~h_{70}$ km s$^{-1}$ Mpc$^{-1}$, $\Omega_{\rm M} = 0.3$, $\Omega_\Lambda = 0.7$. For the studies using the DLA method, we calculated $\Omega_{\rm HI} = \delta_{\rm HI} \Omega_g^{\rm DLA}/\mu$, where $\delta_{\rm HI} = 1.2$ estimates the contribution from systems below DLA threshold and $\mu = 1.3$ accounts for the mass contribution of helium. A gradual increase for $\Omega_{\rm HI}$ with redshift is shown, though with a large scatter around $z \sim 1$. This trend is modelled well by {\sc MUFASA}. The \citet{2015MNRAS.453.2315K} model drops at high redshifts, where there may be significant gas reservoirs in the circumgalactic medium. Our result (red star) agrees with the studies in local Universe, and is consistent with both models.
        
        \begin{figure}
            \includegraphics[width=8cm,height=6cm]{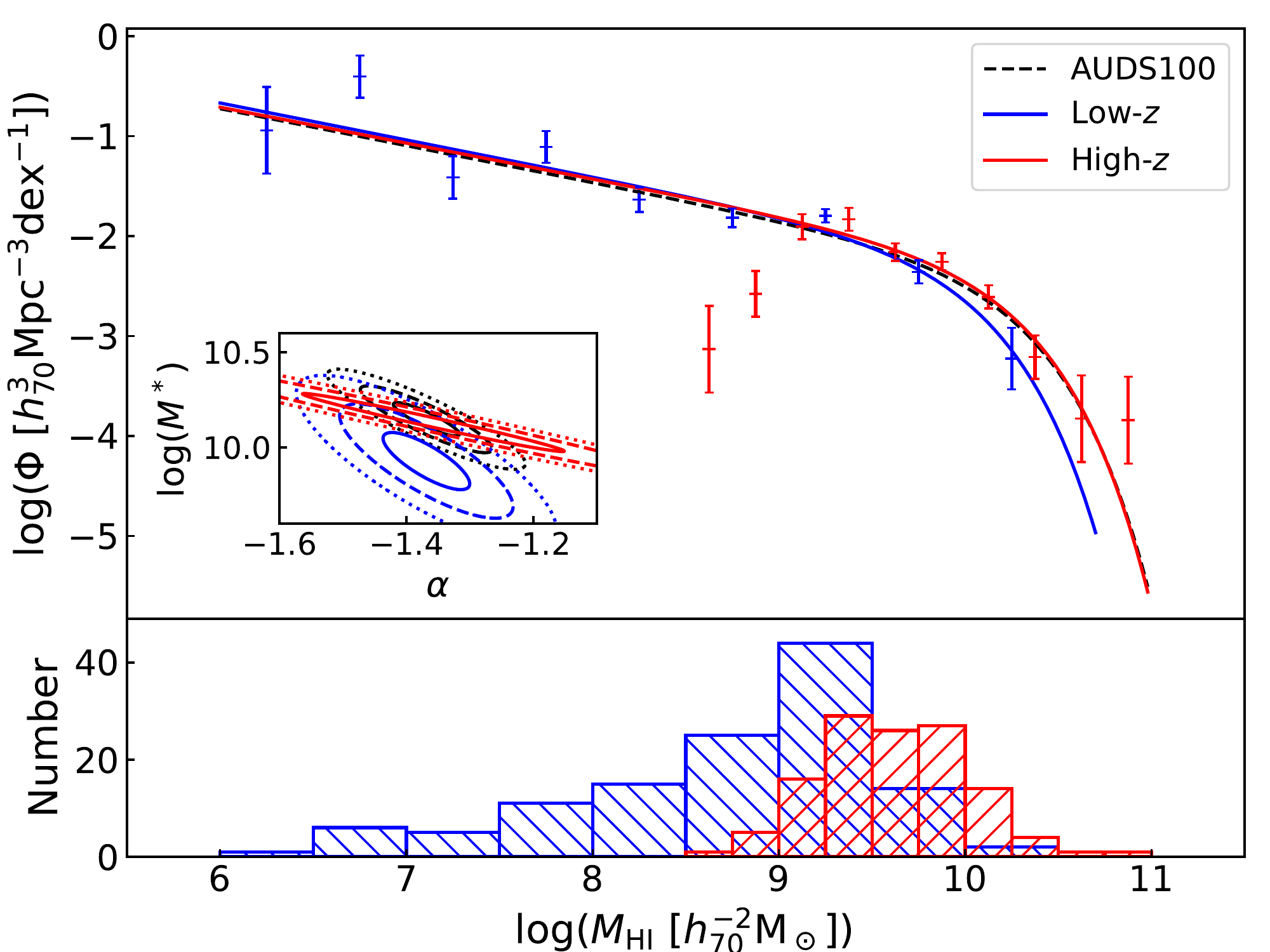}
            \caption{In the lower panel, we show the distribution of galaxy HI mass in  the low-$z$ and high-$z$ samples, divided at $z=0.09$. The red hatched bars represent the high-$z$ sample, while the blue hatched bars represent the low-$z$ sample. The HIMF for both samples is shown in the upper panel in same color. The HIMF for the AUDS100 sample is shown by the black dashed line. The error ellipse in $\alpha$-$\log(M^*)$ plane is also shown in the upper panel, where $M^*$ is in units of $h_{70}^{-2}{\rm M}_\odot$. The bin width is 0.5 dex in the low-$z$ sample, and 0.25 dex in the high-$z$ sample.}
            \centering\label{Fig_20}
        \end{figure}
        
	    In order to investigate whether the HIMF can evolve despite the lack of any apparent change in $\Omega_{\rm HI}$, we divide the AUDS100 sample into two equal sub-samples at $z=0.09$. There are 123 galaxies in low-redshift sample and 124 in the high-redshift sample. The mass distribution in each sample is shown in Figure \ref{Fig_20}. We compute the HIMF using the $\Sigma 1/V_{\rm max}$ (MML) method, and show the results in Figure \ref{Fig_20} and Table \ref{Tab_04}. Due to the large deviation in the HIMF for the high-redshift sample in the lowest two bins, we removed the three lowest-mass galaxies solely for the purpose of fitting the Schechter function. Both the low and high-redshift samples have slopes consistent with the AUDS100 sample within the $1\sigma$ uncertainty, although the high-redshift sample misses the low-mass galaxies required to better constrain this. The high-redshift sample appears to have a larger characteristic mass. However, there are only two galaxies with $\log(M_{\rm HI}~[h_{70}^{-2}{\rm M}_\odot])>10.0$ in the low-redshift sample. The formal significance of the difference is only $1\sigma$. To mitigate the effect of the different mass ranges in the two sub-samples shown in Figure \ref{Fig_20}, we choose a common range ($10^{8.94}$ -- $10^{10.40}$ $h_{70}^{-2} {\rm M}_\odot$). In this range, there are 62 galaxies in the low-$z$ sample and 119 galaxies in the high-$z$ sample. The small numbers give large uncertainties for both $M^*$ and $\alpha$ in the low-$z$ sample, but the significance of the difference between the high- and low-$z$ sample reduces further. However, as already noted, there remains a $2\sigma$ difference when comparing the mean AUDS100 results with the larger HIPASS and ALFALFA samples (see Table \ref{Tab_05}). The MeerKAT LADUMA survey \citep{2016mks..confE...4B, 2018AAS...23123107B} will detect more galaxies with redshift up to $z \sim 1$, which allows us to explore further on the evolution of HIMF.
	    
\section{Conclusions}\label{Conclusions}
    
    We employed the Arecibo Telescope to perform the Arecibo Ultra-Deep Survey (AUDS), the deepest blind HI survey so far conducted with this telescope. The survey area is 1.35 deg$^2$ with redshift coverage up to 0.16 (the limit of the ALFA receiver). The total on-source integration time exceeds 700 hours, and the final noise level is $\sim$75 $\mu$Jy at a frequency resolution of 21.4 kHz. Our final AUDS100 catalogue contains 247 galaxies with an HI mass range from $10^{6.32}$ to $10^{10.76}$ $h_{70}^{-2}{\rm M}_\odot$. For statistical purposes, a high-completeness, high-reliability sub-sample, consisting of 148 galaxies was also created. The AUDS100 catalogue consists galaxies detected by a source-finder and by manual inspection of the data cubes. Simulations were performed to measure completeness, and checked using  a modified $V/V_{\rm max}$ test. 
    
    The HI mass function (HIMF) was computed using the $\Sigma 1/V_{\rm max}$ and 2D stepwise maximum-likelihood (2DSWML) methods. Due to the small field-of-view, cosmic variance in AUDS is substantial, but was able to be corrected using the measured density of SDSS spectroscopic data for blue galaxies which have a similar density bias to HI-rich galaxies. Fits to the HIMF were made using different methods, including a modified maximum likelihood method, which makes better allowance for errors and limits, and is robust against binning artefacts. The best-fit parameters for a Schechter function fit are: $\Phi_* = 2.41 (0.57) \times 10^{-3} h_{70}^3 {\rm Mpc}^{-3}$, $\log(M^*[h_{70}^{-2}{\rm M}_\odot]) = 10.15 (0.09)$, and $\alpha = -1.37(0.05)$. Using these parameters, we compute the AUDS cosmic HI density, $\Omega_{\rm HI} = 3.55(0.30) \times 10^{-4} h_{70}^{-1}$. Our result is consistent with previous measurements from HIPASS and ALFALFA.
    
    Evolutionary trends were inspected by dividing the AUDS100 catalogue into low and high-redshift sub-samples. A slight increase in the characteristic mass $M^*$ is apparent in the higher redshift bin, which becomes more significant when compared with HIPASS and ALFALFA. No evolutionary trend was detected for $\Omega_{\rm HI}$, in line with other recent studies.
    
    Finally, we investigated environmental trends by separately computing HIMFs in the two AUDS fields which have galaxy number densities differ by a factor of two. However, no clear changes were measured between the two fields.
    
    Further exploration of the effect of redshift on the HIMF and $\Omega_{\rm HI}$ will best be performed in future with deep blind HI surveys using ASKAP, FAST and MeerKAT and, eventually, the Square Kilometre Array (SKA). Stacking and intensity mapping experiments have additional dependencies on stronger evolutionary trends at optical/IR wavebands. Further exploration of the effect of environment will be best explored with wide-area blind HI surveys using ASKAP (WALLABY; \citet{2020Ap&SS.365..118K}), FAST (CRAFTS; \citet{8958208}) and WSRT/APERTIF (WEAVE-Apertif Survey; \citet{2020AAS...23545903H}).
    
\section*{Acknowledgements}

    We would like to thank the China Scholarship Council (CSC) from the Ministry of Education of P.R. China and the International Centre for Radio Astronomy Research (ICRAR) in the University of Western Australia (UWA) for financial support during Hongwei Xi studying at ICRAR. The Arecibo Observatory is operated by SRI International under a cooperative agreement with the National Science Foundation (AST-1100968), and in alliance with Ana G. Méndez-Universidad Metropolitana, and the Universities Space Research Association. In light of the recent loss of the telescope, we also wish to thank the staff of the Arecibo Observatory for their tremendous support during the planning and execution of these observations, and for the careful curation and shipping of the data. Without their support and the far-reaching capability of this telescope, this research would not have been possible. Parts of this research were supported by the Australian Research Council Centre of Excellence for All Sky Astrophysics in 3 Dimensions (ASTRO 3D), through project number CE170100013. This work was supported by resources provided by the Pawsey Supercomputing Centre with funding from the Australian Government and the Government of Western Australia.

\section*{Data availability}

    Readers interested in the raw data can contact Arecibo Observatory. The derived data supporting the findings of this study are available from the corresponding author on reasonable request.
    



\bibliographystyle{mnras}
\bibliography{References} 

\begin{thebibliography}{}
\makeatletter
\relax
\def\mn@urlcharsother{\let\do\@makeother \do\$\do\&\do\#\do\^\do\_\do\%\do\~}
\def\mn@doi{\begingroup\mn@urlcharsother \@ifnextchar [ {\mn@doi@}
  {\mn@doi@[]}}
\def\mn@doi@[#1]#2{\def\@tempa{#1}\ifx\@tempa\@empty \href
  {http://dx.doi.org/#2} {doi:#2}\else \href {http://dx.doi.org/#2} {#1}\fi
  \endgroup}
\def\mn@eprint#1#2{\mn@eprint@#1:#2::\@nil}
\def\mn@eprint@arXiv#1{\href {http://arxiv.org/abs/#1} {{\tt arXiv:#1}}}
\def\mn@eprint@dblp#1{\href {http://dblp.uni-trier.de/rec/bibtex/#1.xml}
  {dblp:#1}}
\def\mn@eprint@#1:#2:#3:#4\@nil{\def\@tempa {#1}\def\@tempb {#2}\def\@tempc
  {#3}\ifx \@tempc \@empty \let \@tempc \@tempb \let \@tempb \@tempa \fi \ifx
  \@tempb \@empty \def\@tempb {arXiv}\fi \@ifundefined
  {mn@eprint@\@tempb}{\@tempb:\@tempc}{\expandafter \expandafter \csname
  mn@eprint@\@tempb\endcsname \expandafter{\@tempc}}}

\bibitem[\protect\citeauthoryear{{Abolfathi} et~al.,}{{Abolfathi}
  et~al.}{2018}]{2018ApJS..235...42A}
{Abolfathi} B.,  et~al., 2018, \mn@doi [\apjs] {10.3847/1538-4365/aa9e8a},
  \href {https://ui.adsabs.harvard.edu/abs/2018ApJS..235...42A} {235, 42}

\bibitem[\protect\citeauthoryear{{Baker}, {Blyth}, {Holwerda}  \& {LADUMA
  Team}}{{Baker} et~al.}{2018}]{2018AAS...23123107B}
{Baker} A.~J.,  {Blyth} S.,  {Holwerda} B.~W.,   {LADUMA Team} 2018, in
  American Astronomical Society Meeting Abstracts \#231. p. 231.07

\bibitem[\protect\citeauthoryear{{Barnes} et~al.,}{{Barnes}
  et~al.}{2001}]{2001MNRAS.322..486B}
{Barnes} D.~G.,  et~al., 2001, \mn@doi [\mnras]
  {10.1046/j.1365-8711.2001.04102.x}, \href
  {https://ui.adsabs.harvard.edu/abs/2001MNRAS.322..486B} {322, 486}

\bibitem[\protect\citeauthoryear{{Bird}, {Garnett}  \& {Ho}}{{Bird}
  et~al.}{2017}]{2017MNRAS.466.2111B}
{Bird} S.,  {Garnett} R.,   {Ho} S.,  2017, \mn@doi [\mnras]
  {10.1093/mnras/stw3246}, \href
  {https://ui.adsabs.harvard.edu/abs/2017MNRAS.466.2111B} {466, 2111}

\bibitem[\protect\citeauthoryear{{Blyth} et~al.,}{{Blyth}
  et~al.}{2016}]{2016mks..confE...4B}
{Blyth} S.,  et~al., 2016, in MeerKAT Science: On the Pathway to the SKA. p.~4

\bibitem[\protect\citeauthoryear{{Braun}}{{Braun}}{2012}]{2012ApJ...749...87B}
{Braun} R.,  2012, \mn@doi [\apj] {10.1088/0004-637X/749/1/87}, \href
  {https://ui.adsabs.harvard.edu/abs/2012ApJ...749...87B} {749, 87}

\bibitem[\protect\citeauthoryear{{Catinella}, {Haynes}, {Giovanelli}, {Gardner}
   \& {Connolly}}{{Catinella} et~al.}{2008}]{2008ApJ...685L..13C}
{Catinella} B.,  {Haynes} M.~P.,  {Giovanelli} R.,  {Gardner} J.~P.,
  {Connolly} A.~J.,  2008, \mn@doi [\apjl] {10.1086/592328}, \href
  {https://ui.adsabs.harvard.edu/abs/2008ApJ...685L..13C} {685, L13}

\bibitem[\protect\citeauthoryear{{Chang}, {Pen}, {Bandura}  \&
  {Peterson}}{{Chang} et~al.}{2010}]{2010Natur.466..463C}
{Chang} T.-C.,  {Pen} U.-L.,  {Bandura} K.,   {Peterson} J.~B.,  2010, \mn@doi
  [\nat] {10.1038/nature09187}, \href
  {https://ui.adsabs.harvard.edu/abs/2010Natur.466..463C} {466, 463}

\bibitem[\protect\citeauthoryear{{Condon}, {Cotton}, {Greisen}, {Yin},
  {Perley}, {Taylor}  \& {Broderick}}{{Condon}
  et~al.}{1998}]{1998AJ....115.1693C}
{Condon} J.~J.,  {Cotton} W.~D.,  {Greisen} E.~W.,  {Yin} Q.~F.,  {Perley}
  R.~A.,  {Taylor} G.~B.,   {Broderick} J.~J.,  1998, \mn@doi [\aj]
  {10.1086/300337}, \href
  {https://ui.adsabs.harvard.edu/abs/1998AJ....115.1693C} {115, 1693}

\bibitem[\protect\citeauthoryear{{Courtois} \& {Tully}}{{Courtois} \&
  {Tully}}{2015}]{2015MNRAS.447.1531C}
{Courtois} H.~M.,  {Tully} R.~B.,  2015, \mn@doi [\mnras]
  {10.1093/mnras/stu2405}, \href
  {https://ui.adsabs.harvard.edu/abs/2015MNRAS.447.1531C} {447, 1531}

\bibitem[\protect\citeauthoryear{{Crighton} et~al.,}{{Crighton}
  et~al.}{2015}]{2015MNRAS.452..217C}
{Crighton} N. H.~M.,  et~al., 2015, \mn@doi [\mnras] {10.1093/mnras/stv1182},
  \href {https://ui.adsabs.harvard.edu/abs/2015MNRAS.452..217C} {452, 217}

\bibitem[\protect\citeauthoryear{{Dalcanton}, {Spergel}  \&
  {Summers}}{{Dalcanton} et~al.}{1997}]{1997ApJ...482..659D}
{Dalcanton} J.~J.,  {Spergel} D.~N.,   {Summers} F.~J.,  1997, \mn@doi [\apj]
  {10.1086/304182}, \href
  {https://ui.adsabs.harvard.edu/abs/1997ApJ...482..659D} {482, 659}

\bibitem[\protect\citeauthoryear{{Dav{\'e}}, {Rafieferantsoa}, {Thompson}  \&
  {Hopkins}}{{Dav{\'e}} et~al.}{2017}]{2017MNRAS.467..115D}
{Dav{\'e}} R.,  {Rafieferantsoa} M.~H.,  {Thompson} R.~J.,   {Hopkins} P.~F.,
  2017, \mn@doi [\mnras] {10.1093/mnras/stx108}, \href
  {https://ui.adsabs.harvard.edu/abs/2017MNRAS.467..115D} {467, 115}

\bibitem[\protect\citeauthoryear{{Davis} \& {Huchra}}{{Davis} \&
  {Huchra}}{1982}]{1982ApJ...254..437D}
{Davis} M.,  {Huchra} J.,  1982, \mn@doi [\apj] {10.1086/159751}, \href
  {https://ui.adsabs.harvard.edu/abs/1982ApJ...254..437D} {254, 437}

\bibitem[\protect\citeauthoryear{{Delhaize}, {Meyer}, {Staveley-Smith}  \&
  {Boyle}}{{Delhaize} et~al.}{2013}]{2013MNRAS.433.1398D}
{Delhaize} J.,  {Meyer} M.~J.,  {Staveley-Smith} L.,   {Boyle} B.~J.,  2013,
  \mn@doi [\mnras] {10.1093/mnras/stt810}, \href
  {https://ui.adsabs.harvard.edu/abs/2013MNRAS.433.1398D} {433, 1398}

\bibitem[\protect\citeauthoryear{{Donley} et~al.,}{{Donley}
  et~al.}{2005}]{2005AJ....129..220D}
{Donley} J.~L.,  et~al., 2005, \mn@doi [\aj] {10.1086/426320}, \href
  {https://ui.adsabs.harvard.edu/abs/2005AJ....129..220D} {129, 220}

\bibitem[\protect\citeauthoryear{{Doyle} et~al.,}{{Doyle}
  et~al.}{2005}]{2005MNRAS.361...34D}
{Doyle} M.~T.,  et~al., 2005, \mn@doi [\mnras]
  {10.1111/j.1365-2966.2005.09159.x}, \href
  {https://ui.adsabs.harvard.edu/abs/2005MNRAS.361...34D} {361, 34}

\bibitem[\protect\citeauthoryear{{Driver} \& {Robotham}}{{Driver} \&
  {Robotham}}{2010}]{2010MNRAS.407.2131D}
{Driver} S.~P.,  {Robotham} A. S.~G.,  2010, \mn@doi [\mnras]
  {10.1111/j.1365-2966.2010.17028.x}, \href
  {https://ui.adsabs.harvard.edu/abs/2010MNRAS.407.2131D} {407, 2131}

\bibitem[\protect\citeauthoryear{{Efstathiou}, {Ellis}  \&
  {Peterson}}{{Efstathiou} et~al.}{1988}]{1988MNRAS.232..431E}
{Efstathiou} G.,  {Ellis} R.~S.,   {Peterson} B.~A.,  1988, \mn@doi [\mnras]
  {10.1093/mnras/232.2.431}, \href
  {https://ui.adsabs.harvard.edu/abs/1988MNRAS.232..431E} {232, 431}

\bibitem[\protect\citeauthoryear{{Freudling} et~al.,}{{Freudling}
  et~al.}{2011}]{2011ApJ...727...40F}
{Freudling} W.,  et~al., 2011, \mn@doi [\apj] {10.1088/0004-637X/727/1/40},
  \href {https://ui.adsabs.harvard.edu/abs/2011ApJ...727...40F} {727, 40}

\bibitem[\protect\citeauthoryear{{Giovanelli} et~al.,}{{Giovanelli}
  et~al.}{2005}]{2005AJ....130.2598G}
{Giovanelli} R.,  et~al., 2005, \mn@doi [\aj] {10.1086/497431}, \href
  {https://ui.adsabs.harvard.edu/abs/2005AJ....130.2598G} {130, 2598}

\bibitem[\protect\citeauthoryear{{Grasha}, {Darling}, {Leroy}  \&
  {Bolatto}}{{Grasha} et~al.}{2020}]{2020MNRAS.498..883G}
{Grasha} K.,  {Darling} J.,  {Leroy} A.~K.,   {Bolatto} A.~D.,  2020, \mn@doi
  [\mnras] {10.1093/mnras/staa2521}, \href
  {https://ui.adsabs.harvard.edu/abs/2020MNRAS.498..883G} {498, 883}

\bibitem[\protect\citeauthoryear{{Hess}, {Falcon-Barroso}, {Ascasibar},
  {Perez-Martin}, {Serra}, {Weijmans}  \& {Weave-Apertif Team}}{{Hess}
  et~al.}{2020}]{2020AAS...23545903H}
{Hess} K.~M.,  {Falcon-Barroso} J.,  {Ascasibar} Y.,  {Perez-Martin} I.,
  {Serra} P.,  {Weijmans} A.,   {Weave-Apertif Team} 2020, in American
  Astronomical Society Meeting Abstracts \#235. p. 459.03

\bibitem[\protect\citeauthoryear{{Hoppmann}}{{Hoppmann}}{2014}]{2014UWAThesisH}
{Hoppmann} L.,  2014, PhD thesis, University of Western Australia

\bibitem[\protect\citeauthoryear{{Hoppmann}, {Staveley-Smith}, {Freudling},
  {Zwaan}, {Minchin}  \& {Calabretta}}{{Hoppmann}
  et~al.}{2015}]{2015MNRAS.452.3726H}
{Hoppmann} L.,  {Staveley-Smith} L.,  {Freudling} W.,  {Zwaan} M.~A.,
  {Minchin} R.~F.,   {Calabretta} M.~R.,  2015, \mn@doi [\mnras]
  {10.1093/mnras/stv1084}, \href
  {https://ui.adsabs.harvard.edu/abs/2015MNRAS.452.3726H} {452, 3726}

\bibitem[\protect\citeauthoryear{{Hu} et~al.,}{{Hu}
  et~al.}{2019}]{2019MNRAS.489.1619H}
{Hu} W.,  et~al., 2019, \mn@doi [\mnras] {10.1093/mnras/stz2038}, \href
  {https://ui.adsabs.harvard.edu/abs/2019MNRAS.489.1619H} {489, 1619}

\bibitem[\protect\citeauthoryear{{Jones}, {Papastergis}, {Haynes}  \&
  {Giovanelli}}{{Jones} et~al.}{2016}]{2016MNRAS.457.4393J}
{Jones} M.~G.,  {Papastergis} E.,  {Haynes} M.~P.,   {Giovanelli} R.,  2016,
  \mn@doi [\mnras] {10.1093/mnras/stw263}, \href
  {https://ui.adsabs.harvard.edu/abs/2016MNRAS.457.4393J} {457, 4393}

\bibitem[\protect\citeauthoryear{{Jones}, {Haynes}, {Giovanelli}  \&
  {Moorman}}{{Jones} et~al.}{2018}]{2018MNRAS.477....2J}
{Jones} M.~G.,  {Haynes} M.~P.,  {Giovanelli} R.,   {Moorman} C.,  2018,
  \mn@doi [\mnras] {10.1093/mnras/sty521}, \href
  {https://ui.adsabs.harvard.edu/abs/2018MNRAS.477....2J} {477, 2}

\bibitem[\protect\citeauthoryear{{Kanekar}, {Sethi}  \&
  {Dwarakanath}}{{Kanekar} et~al.}{2016}]{2016ApJ...818L..28K}
{Kanekar} N.,  {Sethi} S.,   {Dwarakanath} K.~S.,  2016, \mn@doi [\apjl]
  {10.3847/2041-8205/818/2/L28}, \href
  {https://ui.adsabs.harvard.edu/abs/2016ApJ...818L..28K} {818, L28}

\bibitem[\protect\citeauthoryear{{Kim}, {Wyithe}, {Power}, {Park}, {Lagos}  \&
  {Baugh}}{{Kim} et~al.}{2015}]{2015MNRAS.453.2315K}
{Kim} H.-S.,  {Wyithe} J.~S.~B.,  {Power} C.,  {Park} J.,  {Lagos} C. d.~P.,
  {Baugh} C.~M.,  2015, \mn@doi [\mnras] {10.1093/mnras/stv1822}, \href
  {https://ui.adsabs.harvard.edu/abs/2015MNRAS.453.2315K} {453, 2315}

\bibitem[\protect\citeauthoryear{{Koribalski} et~al.,}{{Koribalski}
  et~al.}{2020}]{2020Ap&SS.365..118K}
{Koribalski} B.~S.,  et~al., 2020, \mn@doi [\apss]
  {10.1007/s10509-020-03831-4}, \href
  {https://ui.adsabs.harvard.edu/abs/2020Ap&SS.365..118K} {365, 118}

\bibitem[\protect\citeauthoryear{{Lagos}, {Baugh}, {Zwaan}, {Lacey},
  {Gonzalez-Perez}, {Power}, {Swinbank}  \& {van Kampen}}{{Lagos}
  et~al.}{2014}]{2014MNRAS.440..920L}
{Lagos} C.~D.~P.,  {Baugh} C.~M.,  {Zwaan} M.~A.,  {Lacey} C.~G.,
  {Gonzalez-Perez} V.,  {Power} C.,  {Swinbank} A.~M.,   {van Kampen} E.,
  2014, \mn@doi [\mnras] {10.1093/mnras/stu266}, \href
  {https://ui.adsabs.harvard.edu/abs/2014MNRAS.440..920L} {440, 920}

\bibitem[\protect\citeauthoryear{{Lah} et~al.,}{{Lah}
  et~al.}{2007}]{2007MNRAS.376.1357L}
{Lah} P.,  et~al., 2007, \mn@doi [\mnras] {10.1111/j.1365-2966.2007.11540.x},
  \href {https://ui.adsabs.harvard.edu/abs/2007MNRAS.376.1357L} {376, 1357}

\bibitem[\protect\citeauthoryear{{Li} \& {Duan}}{{Li} \&
  {Duan}}{2019}]{8958208}
{Li} D.,  {Duan} R.,  2019, in 2019 IEEE International Conference on
  Microwaves, Antennas, Communications and Electronic Systems (COMCAS). pp~1--4

\bibitem[\protect\citeauthoryear{{Loveday}}{{Loveday}}{2000}]{2000MNRAS.312..557L}
{Loveday} J.,  2000, \mn@doi [\mnras] {10.1046/j.1365-8711.2000.03179.x}, \href
  {https://ui.adsabs.harvard.edu/abs/2000MNRAS.312..557L} {312, 557}

\bibitem[\protect\citeauthoryear{{Martin}, {Papastergis}, {Giovanelli},
  {Haynes}, {Springob}  \& {Stierwalt}}{{Martin}
  et~al.}{2010}]{2010ApJ...723.1359M}
{Martin} A.~M.,  {Papastergis} E.,  {Giovanelli} R.,  {Haynes} M.~P.,
  {Springob} C.~M.,   {Stierwalt} S.,  2010, \mn@doi [\apj]
  {10.1088/0004-637X/723/2/1359}, \href
  {https://ui.adsabs.harvard.edu/abs/2010ApJ...723.1359M} {723, 1359}

\bibitem[\protect\citeauthoryear{{Masui} et~al.,}{{Masui}
  et~al.}{2013}]{2013ApJ...763L..20M}
{Masui} K.~W.,  et~al., 2013, \mn@doi [\apjl] {10.1088/2041-8205/763/1/L20},
  \href {https://ui.adsabs.harvard.edu/abs/2013ApJ...763L..20M} {763, L20}

\bibitem[\protect\citeauthoryear{{McGaugh}}{{McGaugh}}{1996}]{1996MNRAS.280..337M}
{McGaugh} S.~S.,  1996, \mn@doi [\mnras] {10.1093/mnras/280.2.337}, \href
  {https://ui.adsabs.harvard.edu/abs/1996MNRAS.280..337M} {280, 337}

\bibitem[\protect\citeauthoryear{{Meyer} et~al.,}{{Meyer}
  et~al.}{2004}]{2004MNRAS.350.1195M}
{Meyer} M.~J.,  et~al., 2004, \mn@doi [\mnras]
  {10.1111/j.1365-2966.2004.07710.x}, \href
  {https://ui.adsabs.harvard.edu/abs/2004MNRAS.350.1195M} {350, 1195}

\bibitem[\protect\citeauthoryear{{Meyer}, {Robotham}, {Obreschkow},
  {Westmeier}, {Duffy}  \& {Staveley-Smith}}{{Meyer}
  et~al.}{2017}]{2017PASA...34...52M}
{Meyer} M.,  {Robotham} A.,  {Obreschkow} D.,  {Westmeier} T.,  {Duffy} A.~R.,
   {Staveley-Smith} L.,  2017, \mn@doi [\pasa] {10.1017/pasa.2017.31}, \href
  {https://ui.adsabs.harvard.edu/abs/2017PASA...34...52M} {34, 52}

\bibitem[\protect\citeauthoryear{{Moorman}, {Vogeley}, {Hoyle}, {Pan}, {Haynes}
   \& {Giovanelli}}{{Moorman} et~al.}{2014}]{2014MNRAS.444.3559M}
{Moorman} C.~M.,  {Vogeley} M.~S.,  {Hoyle} F.,  {Pan} D.~C.,  {Haynes} M.~P.,
   {Giovanelli} R.,  2014, \mn@doi [\mnras] {10.1093/mnras/stu1674}, \href
  {https://ui.adsabs.harvard.edu/abs/2014MNRAS.444.3559M} {444, 3559}

\bibitem[\protect\citeauthoryear{{Neeleman}, {Prochaska}, {Ribaudo}, {Lehner},
  {Howk}, {Rafelski}  \& {Kanekar}}{{Neeleman}
  et~al.}{2016}]{2016ApJ...818..113N}
{Neeleman} M.,  {Prochaska} J.~X.,  {Ribaudo} J.,  {Lehner} N.,  {Howk} J.~C.,
  {Rafelski} M.,   {Kanekar} N.,  2016, \mn@doi [\apj]
  {10.3847/0004-637X/818/2/113}, \href
  {https://ui.adsabs.harvard.edu/abs/2016ApJ...818..113N} {818, 113}

\bibitem[\protect\citeauthoryear{{Noterdaeme}, {Petitjean}, {Ledoux}  \&
  {Srianand}}{{Noterdaeme} et~al.}{2009}]{2009A&A...505.1087N}
{Noterdaeme} P.,  {Petitjean} P.,  {Ledoux} C.,   {Srianand} R.,  2009, \mn@doi
  [\aap] {10.1051/0004-6361/200912768}, \href
  {https://ui.adsabs.harvard.edu/abs/2009A&A...505.1087N} {505, 1087}

\bibitem[\protect\citeauthoryear{{Noterdaeme} et~al.,}{{Noterdaeme}
  et~al.}{2012}]{2012A&A...547L...1N}
{Noterdaeme} P.,  et~al., 2012, \mn@doi [\aap] {10.1051/0004-6361/201220259},
  \href {https://ui.adsabs.harvard.edu/abs/2012A&A...547L...1N} {547, L1}

\bibitem[\protect\citeauthoryear{{Obreschkow}, {Murray}, {Robotham}  \&
  {Westmeier}}{{Obreschkow} et~al.}{2018}]{2018MNRAS.474.5500O}
{Obreschkow} D.,  {Murray} S.~G.,  {Robotham} A.~S.~G.,   {Westmeier} T.,
  2018, \mn@doi [\mnras] {10.1093/mnras/stx3155}, \href
  {https://ui.adsabs.harvard.edu/abs/2018MNRAS.474.5500O} {474, 5500}

\bibitem[\protect\citeauthoryear{{P{\'e}roux}, {McMahon}, {Storrie-Lombardi}
  \& {Irwin}}{{P{\'e}roux} et~al.}{2003}]{2003MNRAS.346.1103P}
{P{\'e}roux} C.,  {McMahon} R.~G.,  {Storrie-Lombardi} L.~J.,   {Irwin} M.~J.,
  2003, \mn@doi [\mnras] {10.1111/j.1365-2966.2003.07129.x}, \href
  {https://ui.adsabs.harvard.edu/abs/2003MNRAS.346.1103P} {346, 1103}

\bibitem[\protect\citeauthoryear{{Prochaska} \& {Wolfe}}{{Prochaska} \&
  {Wolfe}}{2009}]{2009ApJ...696.1543P}
{Prochaska} J.~X.,  {Wolfe} A.~M.,  2009, \mn@doi [\apj]
  {10.1088/0004-637X/696/2/1543}, \href
  {https://ui.adsabs.harvard.edu/abs/2009ApJ...696.1543P} {696, 1543}

\bibitem[\protect\citeauthoryear{{Prochaska}, {Herbert-Fort}  \&
  {Wolfe}}{{Prochaska} et~al.}{2005}]{2005ApJ...635..123P}
{Prochaska} J.~X.,  {Herbert-Fort} S.,   {Wolfe} A.~M.,  2005, \mn@doi [\apj]
  {10.1086/497287}, \href
  {https://ui.adsabs.harvard.edu/abs/2005ApJ...635..123P} {635, 123}

\bibitem[\protect\citeauthoryear{{Rao} \& {Turnshek}}{{Rao} \&
  {Turnshek}}{2000}]{2000ApJS..130....1R}
{Rao} S.~M.,  {Turnshek} D.~A.,  2000, \mn@doi [\apjs] {10.1086/317344}, \href
  {https://ui.adsabs.harvard.edu/abs/2000ApJS..130....1R} {130, 1}

\bibitem[\protect\citeauthoryear{{Rao}, {Turnshek}  \& {Nestor}}{{Rao}
  et~al.}{2006}]{2006ApJ...636..610R}
{Rao} S.~M.,  {Turnshek} D.~A.,   {Nestor} D.~B.,  2006, \mn@doi [\apj]
  {10.1086/498132}, \href
  {https://ui.adsabs.harvard.edu/abs/2006ApJ...636..610R} {636, 610}

\bibitem[\protect\citeauthoryear{{Rao}, {Turnshek}, {Sardane}  \&
  {Monier}}{{Rao} et~al.}{2017}]{2017MNRAS.471.3428R}
{Rao} S.~M.,  {Turnshek} D.~A.,  {Sardane} G.~M.,   {Monier} E.~M.,  2017,
  \mn@doi [\mnras] {10.1093/mnras/stx1787}, \href
  {https://ui.adsabs.harvard.edu/abs/2017MNRAS.471.3428R} {471, 3428}

\bibitem[\protect\citeauthoryear{{Rauzy}}{{Rauzy}}{2001}]{2001MNRAS.324...51R}
{Rauzy} S.,  2001, \mn@doi [\mnras] {10.1046/j.1365-8711.2001.04078.x}, \href
  {https://ui.adsabs.harvard.edu/abs/2001MNRAS.324...51R} {324, 51}

\bibitem[\protect\citeauthoryear{{Reynolds} et~al.,}{{Reynolds}
  et~al.}{2019}]{2019MNRAS.482.3591R}
{Reynolds} T.~N.,  et~al., 2019, \mn@doi [\mnras] {10.1093/mnras/sty2930},
  \href {https://ui.adsabs.harvard.edu/abs/2019MNRAS.482.3591R} {482, 3591}

\bibitem[\protect\citeauthoryear{{Rhee}, {Zwaan}, {Briggs}, {Chengalur}, {Lah},
  {Oosterloo}  \& {van der Hulst}}{{Rhee} et~al.}{2013}]{2013MNRAS.435.2693R}
{Rhee} J.,  {Zwaan} M.~A.,  {Briggs} F.~H.,  {Chengalur} J.~N.,  {Lah} P.,
  {Oosterloo} T.,   {van der Hulst} T.,  2013, \mn@doi [\mnras]
  {10.1093/mnras/stt1481}, \href
  {https://ui.adsabs.harvard.edu/abs/2013MNRAS.435.2693R} {435, 2693}

\bibitem[\protect\citeauthoryear{{Rhee}, {Lah}, {Chengalur}, {Briggs}  \&
  {Colless}}{{Rhee} et~al.}{2016}]{2016MNRAS.460.2675R}
{Rhee} J.,  {Lah} P.,  {Chengalur} J.~N.,  {Briggs} F.~H.,   {Colless} M.,
  2016, \mn@doi [\mnras] {10.1093/mnras/stw1097}, \href
  {https://ui.adsabs.harvard.edu/abs/2016MNRAS.460.2675R} {460, 2675}

\bibitem[\protect\citeauthoryear{{Rhee}, {Lah}, {Briggs}, {Chengalur},
  {Colless}, {Willner}, {Ashby}  \& {Le F{\`e}vre}}{{Rhee}
  et~al.}{2018}]{2018MNRAS.473.1879R}
{Rhee} J.,  {Lah} P.,  {Briggs} F.~H.,  {Chengalur} J.~N.,  {Colless} M.,
  {Willner} S.~P.,  {Ashby} M. L.~N.,   {Le F{\`e}vre} O.,  2018, \mn@doi
  [\mnras] {10.1093/mnras/stx2461}, \href
  {https://ui.adsabs.harvard.edu/abs/2018MNRAS.473.1879R} {473, 1879}

\bibitem[\protect\citeauthoryear{{Said}, {Kraan-Korteweg}  \&
  {Staveley-Smith}}{{Said} et~al.}{2019}]{2019MNRAS.486.1796S}
{Said} K.,  {Kraan-Korteweg} R.~C.,   {Staveley-Smith} L.,  2019, \mn@doi
  [\mnras] {10.1093/mnras/stz956}, \href
  {https://ui.adsabs.harvard.edu/abs/2019MNRAS.486.1796S} {486, 1796}

\bibitem[\protect\citeauthoryear{{Sancisi}, {Fraternali}, {Oosterloo}  \& {van
  der Hulst}}{{Sancisi} et~al.}{2008}]{2008A&ARv..15..189S}
{Sancisi} R.,  {Fraternali} F.,  {Oosterloo} T.,   {van der Hulst} T.,  2008,
  \mn@doi [\aapr] {10.1007/s00159-008-0010-0}, \href
  {https://ui.adsabs.harvard.edu/abs/2008A&ARv..15..189S} {15, 189}

\bibitem[\protect\citeauthoryear{{Schechter}}{{Schechter}}{1976}]{1976ApJ...203..297S}
{Schechter} P.,  1976, \mn@doi [\apj] {10.1086/154079}, \href
  {https://ui.adsabs.harvard.edu/abs/1976ApJ...203..297S} {203, 297}

\bibitem[\protect\citeauthoryear{{Schmidt}}{{Schmidt}}{1968}]{1968ApJ...151..393S}
{Schmidt} M.,  1968, \mn@doi [\apj] {10.1086/149446}, \href
  {https://ui.adsabs.harvard.edu/abs/1968ApJ...151..393S} {151, 393}

\bibitem[\protect\citeauthoryear{{Serra} et~al.,}{{Serra}
  et~al.}{2015a}]{2015MNRAS.448.1922S}
{Serra} P.,  et~al., 2015a, \mn@doi [\mnras] {10.1093/mnras/stv079}, \href
  {https://ui.adsabs.harvard.edu/abs/2015MNRAS.448.1922S} {448, 1922}

\bibitem[\protect\citeauthoryear{{Serra} et~al.,}{{Serra}
  et~al.}{2015b}]{2015MNRAS.452.2680S}
{Serra} P.,  et~al., 2015b, \mn@doi [\mnras] {10.1093/mnras/stv1326}, \href
  {https://ui.adsabs.harvard.edu/abs/2015MNRAS.452.2680S} {452, 2680}

\bibitem[\protect\citeauthoryear{{Solanes}, {Giovanelli}  \&
  {Haynes}}{{Solanes} et~al.}{1996}]{1996ApJ...461..609S}
{Solanes} J.~M.,  {Giovanelli} R.,   {Haynes} M.~P.,  1996, \mn@doi [\apj]
  {10.1086/177089}, \href
  {https://ui.adsabs.harvard.edu/abs/1996ApJ...461..609S} {461, 609}

\bibitem[\protect\citeauthoryear{{Songaila} \& {Cowie}}{{Songaila} \&
  {Cowie}}{2010}]{2010ApJ...721.1448S}
{Songaila} A.,  {Cowie} L.~L.,  2010, \mn@doi [\apj]
  {10.1088/0004-637X/721/2/1448}, \href
  {https://ui.adsabs.harvard.edu/abs/2010ApJ...721.1448S} {721, 1448}

\bibitem[\protect\citeauthoryear{{Sprayberry}, {Impey}, {Irwin}  \&
  {Bothun}}{{Sprayberry} et~al.}{1997}]{1997ApJ...482..104S}
{Sprayberry} D.,  {Impey} C.~D.,  {Irwin} M.~J.,   {Bothun} G.~D.,  1997,
  \mn@doi [\apj] {10.1086/304126}, \href
  {https://ui.adsabs.harvard.edu/abs/1997ApJ...482..104S} {482, 104}

\bibitem[\protect\citeauthoryear{{Springob}, {Haynes}  \&
  {Giovanelli}}{{Springob} et~al.}{2005}]{2005ApJ...621..215S}
{Springob} C.~M.,  {Haynes} M.~P.,   {Giovanelli} R.,  2005, \mn@doi [\apj]
  {10.1086/427432}, \href
  {https://ui.adsabs.harvard.edu/abs/2005ApJ...621..215S} {621, 215}

\bibitem[\protect\citeauthoryear{{Staveley-Smith}, {Kraan-Korteweg},
  {Schr{\"o}der}, {Henning}, {Koribalski}, {Stewart}  \&
  {Heald}}{{Staveley-Smith} et~al.}{2016}]{2016AJ....151...52S}
{Staveley-Smith} L.,  {Kraan-Korteweg} R.~C.,  {Schr{\"o}der} A.~C.,  {Henning}
  P.~A.,  {Koribalski} B.~S.,  {Stewart} I.~M.,   {Heald} G.,  2016, \mn@doi
  [\aj] {10.3847/0004-6256/151/3/52}, \href
  {https://ui.adsabs.harvard.edu/abs/2016AJ....151...52S} {151, 52}

\bibitem[\protect\citeauthoryear{{Strateva} et~al.,}{{Strateva}
  et~al.}{2001}]{2001AJ....122.1861S}
{Strateva} I.,  et~al., 2001, \mn@doi [\aj] {10.1086/323301}, \href
  {https://ui.adsabs.harvard.edu/abs/2001AJ....122.1861S} {122, 1861}

\bibitem[\protect\citeauthoryear{{Westmeier}, {Jurek}, {Obreschkow},
  {Koribalski}  \& {Staveley-Smith}}{{Westmeier}
  et~al.}{2014}]{2014MNRAS.438.1176W}
{Westmeier} T.,  {Jurek} R.,  {Obreschkow} D.,  {Koribalski} B.~S.,
  {Staveley-Smith} L.,  2014, \mn@doi [\mnras] {10.1093/mnras/stt2266}, \href
  {https://ui.adsabs.harvard.edu/abs/2014MNRAS.438.1176W} {438, 1176}

\bibitem[\protect\citeauthoryear{{Westmeier} et~al.,}{{Westmeier}
  et~al.}{2017}]{2017MNRAS.472.4832W}
{Westmeier} T.,  et~al., 2017, \mn@doi [\mnras] {10.1093/mnras/stx2289}, \href
  {https://ui.adsabs.harvard.edu/abs/2017MNRAS.472.4832W} {472, 4832}

\bibitem[\protect\citeauthoryear{{Wolfe}, {Gawiser}  \& {Prochaska}}{{Wolfe}
  et~al.}{2005}]{2005ARA&A..43..861W}
{Wolfe} A.~M.,  {Gawiser} E.,   {Prochaska} J.~X.,  2005, \mn@doi [\araa]
  {10.1146/annurev.astro.42.053102.133950}, \href
  {https://ui.adsabs.harvard.edu/abs/2005ARA&A..43..861W} {43, 861}

\bibitem[\protect\citeauthoryear{{Wong} et~al.,}{{Wong}
  et~al.}{2006}]{2006MNRAS.371.1855W}
{Wong} O.~I.,  et~al., 2006, \mn@doi [\mnras]
  {10.1111/j.1365-2966.2006.10846.x}, \href
  {https://ui.adsabs.harvard.edu/abs/2006MNRAS.371.1855W} {371, 1855}

\bibitem[\protect\citeauthoryear{{York} et~al.,}{{York}
  et~al.}{2000}]{2000AJ....120.1579Y}
{York} D.~G.,  et~al., 2000, \mn@doi [\aj] {10.1086/301513}, \href
  {https://ui.adsabs.harvard.edu/abs/2000AJ....120.1579Y} {120, 1579}

\bibitem[\protect\citeauthoryear{{Zafar}, {P{\'e}roux}, {Popping}, {Milliard},
  {Deharveng}  \& {Frank}}{{Zafar} et~al.}{2013}]{2013A&A...556A.141Z}
{Zafar} T.,  {P{\'e}roux} C.,  {Popping} A.,  {Milliard} B.,  {Deharveng}
  J.~M.,   {Frank} S.,  2013, \mn@doi [\aap] {10.1051/0004-6361/201321154},
  \href {https://ui.adsabs.harvard.edu/abs/2013A&A...556A.141Z} {556, A141}

\bibitem[\protect\citeauthoryear{{Zwaan} et~al.,}{{Zwaan}
  et~al.}{2003}]{2003AJ....125.2842Z}
{Zwaan} M.~A.,  et~al., 2003, \mn@doi [\aj] {10.1086/374944}, \href
  {https://ui.adsabs.harvard.edu/abs/2003AJ....125.2842Z} {125, 2842}

\bibitem[\protect\citeauthoryear{{Zwaan} et~al.,}{{Zwaan}
  et~al.}{2004}]{2004MNRAS.350.1210Z}
{Zwaan} M.~A.,  et~al., 2004, \mn@doi [\mnras]
  {10.1111/j.1365-2966.2004.07782.x}, \href
  {https://ui.adsabs.harvard.edu/abs/2004MNRAS.350.1210Z} {350, 1210}

\bibitem[\protect\citeauthoryear{{Zwaan}, {Meyer}, {Staveley-Smith}  \&
  {Webster}}{{Zwaan} et~al.}{2005}]{2005MNRAS.359L..30Z}
{Zwaan} M.~A.,  {Meyer} M.~J.,  {Staveley-Smith} L.,   {Webster} R.~L.,  2005,
  \mn@doi [\mnras] {10.1111/j.1745-3933.2005.00029.x}, \href
  {https://ui.adsabs.harvard.edu/abs/2005MNRAS.359L..30Z} {359, L30}

\makeatother
\end{thebibliography}

\section*{Supporting information}

    Additional Supporting Information can be found in the on- line version of this article.
    The directory ``Spectra'' consists of all the spectra of AUDS100 HI galaxies in .png format. 
    The directory ``Table2'' gives the description file, ``ReadMe'', and data file, ``table2.dat''.


\bsp	
\label{lastpage}
\end{document}